\documentclass[11pt]{article}

\usepackage{amsmath}
\usepackage{amssymb}
\usepackage{slashed}
\usepackage[numbers,sort&compress]{natbib}

\usepackage[letterpaper,margin=1in,bottom=1in]{geometry}
\usepackage{float} 
\usepackage{parskip} 
\usepackage{tabulary} 
\usepackage{color} 
\usepackage{soul} 
\usepackage{subfigure}
\usepackage{graphicx}
\usepackage[section]{placeins} 

\def\snco{stau--neutralino~coannihilation~}
\def\sco{stau--neutralino~coannihilation~}

\def\lhc2{LHC~Run~II}
\def\nsu{nonuniversal~supergravity~models~}
\def\dd{$\cdot\cdot$}

\usepackage{cleveref}

\newcommand{\code}[1]{\texttt{#1}}
\newcommand{\GeV}{~\textrm{GeV}}

\def\co{coannihilation~}

\def\.4{\vspace{-.5cm}}
\newcommand{\ifb}{~\textrm{fb}^{-1}}

\def\beq{\begin{equation}}
\def\be{\begin{equation}}
\def\beqn{\begin{eqnarray}}
\def\ee{\end{equation}}
\def\eeq{\end{equation}}
\def\eeqn{\end{eqnarray}}

\def\co{coannihilation~}

\author{
Amin Aboubrahim\footnote{Email:a.abouibrahim@northeastern.edu},~\,
Pran Nath\footnote{Email:p.nath@northeastern.edu}~\,and
Andrew B. Spisak\footnote{Email:a.spisak@northeastern.edu}
\\~\\
Department of Physics, Northeastern University,
Boston, MA 02115-5000, USA
}

\title{Stau Coannihilation, Compressed Spectrum and SUSY Discovery at the LHC}

\begin{document}
\maketitle
\date

\textbf{Abstract: } 
The lack of observation of supersymmetry thus far implies that the weak supersymmetry scale is larger than what was thought before the LHC era. This observation is strengthened by the Higgs boson mass measurement 
at $\sim 125$ GeV which within  supersymmetric models implies a large loop correction and a weak supersymmetry 
scale lying in the several TeV region. In addition if neutralino is the dark matter, its 
relic density puts further constraints on models often requiring \co to reduce the neutralino relic density to be 
consistent with experimental observation. The \co in turn implies that the mass gap between the LSP and the NLSP
will be small leading to softer final states and making the observation of supersymmetry challenging. 
In this work we investigate stau coannihilation models within supergravity grand unified models and 
the potential of discovery of such models at the LHC in the post Higgs boson discovery era.  We utilize a variety of signal regions to optimize the 
discovery of supersymmetry in the \snco region.
 In the analysis presented we impose the relic density constraint as well as the constraint of the Higgs boson mass. The range of sparticle masses discoverable up to the optimal integrated luminosity of the HL-LHC is investigated. It is found that the mass difference between the stau and the neutralino does not exceed $\sim 20$ GeV over the entire mass range of the models explored.  Thus the discovery of a supersymmetric signal arising from the \snco region will also provide a measurement of the neutralino mass.  The direct detection of neutralino dark matter is  analyzed within the class of \snco models investigated. The analysis is extended to include multi-particle coannihilation where stau along with 
chargino and the second neutralino enter in the \co process.  

\section{Introduction}\label{sec:intro}

Supersymmetry has not been observed thus far, which implies that its scale is higher
than expected before the LHC era. This observation is strengthened by the discovery that the Higgs 
boson~\cite{Englert:1964et, Higgs:1964pj, Guralnik:1964eu} 
mass is $\sim 125$ GeV~\cite{Chatrchyan:2012ufa, Aad:2012tfa}\footnote{For a review of the status of 
supersymmetry after the Higgs boson mass measurement at $\sim 125$ GeV see~\cite{Nath:2015dza}}.
Analysis within high-scale supergravity grand unified model~\cite{msugra} (for a review see~\cite{book}) 
show that the loop correction to the Higgs boson mass in supersymmetry must itself be sizable, which in turn implies a larger value for weak supersymmetry scale lying in the several TeV region~\cite{Akula:2011aa,Arbey:2012dq,susy-higgs,Baer:2015fsa}. There is another constraint that explains the possible reason for the lack of detection of a supersymmetric
 signal. In supergravity grand unified models with R party conservation, neutralino is the LSP over most of the parameter space of models~\cite{Arnowitt:1992aq} 
 and thus a candidate for dark matter. The annihilation of the neutralino in sufficient amounts to have its relic density consistent with the WMAP and then PLANCK experimental results imposes additional constraints. Specifically if the neutralino is Bino like, one needs coannihilation (for early work see~\cite{Griest:1990kh})
 to 
 have consistency with experiment. However, \co implies that the next to lightest supersymmetric particle (the NLSP) must be close
 to the LSP with a small mass gap to ensure efficient annihilation of the LSP. The existence of the small mass gap in turn implies
 that the final states in the decay of the NLSP will be soft making them difficult to detect. Coannihilation appears in supergravity models
 with universal as well as with non-universal boundary conditions at the grand unification scale which lead to 
  a large sparticle landscape~\cite{Feldman:2007zn}.   The large landscape  
 includes non-universalities in the gaugino sector~\cite{Ellis:1985jn,nonuni2} and in the matter and 
Higgs sectors~\cite{NU}. We note in passing that often naturalness criteria are used to argue 
what the scale of weak scale supersymmetry should be. Previously it has been argued that the weak scale could be
large and natural on the hyperbolic branch  
of radiative breaking of the electroweak symmetry~\cite{Chan:1997bi,Feng:1999mn,Chattopadhyay:2003xi,Baer:2003wx,Feldman:2011ud,Akula:2011jx,Ross:2017kjc}.
Additionally, analyses of  naturalness 
including proton stability  from baryon and lepton number violating dimension five operators in 
grand unified theories (for a review of status see~\cite{Nath:2006ut})
along with electroweak symmetry breaking constraints tend to favor the
weak scale of SUSY in the  TeV region~\cite{Liu:2013ula}.

Thus \co necessarily implies that the lightest sparticle spectra which are the prime candidates for detection  are compressed.  
Such compressed  spectra can appear in stau coannihilation, stop coannihilation, gluino coannihilation 
among others (for some recent works on stop coannihilation and gluino \co in the post Higgs boson discovery era see 
\cite{Kaufman:2015nda, Nath:2016kfp}. For recent theory papers related to supersymmetry and compressed
spectrum see~\cite{Dutta:2015exw,Berggren:2015qua,Berggren:2016qjh,LeCompte:2011fh}
and for experimental searches for supersymmetry
with a compressed spectrum see~\cite{Khachatryan:2016mbu,Khachatryan:2016pxa,MORVAJ:2014opa}).
 In this work we extend that analysis to the stau coannihilation region under the Higgs boson mass constraint
 and the relic density constraints.  Stau--neutralino coannihilation has previously been investigated by a number of works~\cite{Berggren:2015qua, Ellis:1998kh,Arnowitt:2008bz,Florez:2016lwi}. Specifically in~\cite{Berggren:2015qua} an analysis has been carried out for the \sco region at \lhc2. However, the analysis of ~\cite{Berggren:2015qua} was limited to neutralino masses below 100 GeV and further the Higgs boson mass constraint and the relic density constraints were not imposed. In this work we use \nsu with nonuniversalities in the gaugino sector to investigate the full range of neutralino and stau masses that are discoverable up to the expected integrated luminosity at the LHC in the future. In our analysis we impose the relic density constraints as well as constraint of the Higgs boson mass. Specifically we use the $\tilde g$SUGRA model~\cite{Akula:2013ioa} where the mass for the $SU(3)_C$ gaugino is much larger than the masses for the 
electroweak gauginos. In this case the universal scalar mass can be rather low lying in the low hundreds of $\GeV$
and it is the gluino which drives the radiative breaking giving much larger masses to the squarks while
the slepton masses remain low. In this case the stop masses can be large enough to give the desired
loop corrections for the Higgs boson mass. Further, the large splitting between the squark masses and the
slepton masses allows for the possibility of a stau-neutralino coannihilation.  
We will also investigate in the $\tilde g$SUGRA framework a multi-particle \co where more than two particles
participate in the \co process. This happens, for example,
 if the neutralino, the stau, and the chargino, and the next to lightest neutralino  
  are clustered together.
In this case one finds more copious set of signatures for discovery. 
 We use signal regions based on those previously published in~\cite{ATLAS:2016uwq, Florez:2016lwi} but optimized for the \snco region. An analysis of dark matter is also given.\\
 
 The outline of the rest of the paper is as follows: In section 2 we discuss how a stau-neutralino \co can arise in a
  high scale model while generating the desired correction to the Higgs boson mass and also satisfying
  the relic density constraints. We also discuss here the possibility of a multiparticle \co involving 
  the neutralino, the stau, the chargino and the second neutralino. 
  In section 3, we discuss the production of supersymmetric particles for the stau and the multiparticle
  \co models. Here we exhibit the  cross sections for the production of the final states
   $\tilde\chi_2^0\tilde\chi_1^\pm$, $\tilde\chi_1^+\tilde\chi_1^-$, $\tilde\tau^+\tilde\tau^-$,
    and $\tilde\tau\tilde\nu_\tau$.  The sparticles in the final states decay with a neutralino and 
     leptons in the final states.   The signature analysis of these requires a knowledge of the backgrounds
     arising from the production and decay of the standard model particles.  Here we use the backgrounds
     published by the SNOWMASS group. 
  Section 4 is devoted to the signature analysis of the high scale models and an
  analysis of the minimum integrated luminosity needed with the LHC operating at $13$ TeV
 for the $5\sigma$ discovery.  Here a comparison of the different signature regions is also 
 made and combined signal region results are exhibited where models are arranged in terms of
 ascending order in the minimum integrated luminosity needed for a $5\sigma$ discovery.
 At the end of this 
 section  we give  an analysis of dark matter cross sections for the models discussed in sections 2-4.
It is shown that the spin independent  neutralino-proton cross sections lie 
 significantly above the neutralino floor and some of the models lie close to the lower bounds
 that will be reached by the next generation direct detection experiments. Conclusions are given 
 in section 5.\\

\section{Stau coannihilation in SUGRA models \label{sec:parameter}}
We have earlier noted that the observation of the Higgs boson mass at $\sim 125$ GeV requires
a large loop correction to its tree value which is below the $Z$-boson mass. The largest correction
arises from the stop masses in the loop and one needs an average stop mass in the several  TeV region. 
 In SUGRA models with universal soft parameters at the grand unification scale,   this would indicate
 a large universal scalar mass $m_0$ if we wish to have the charginos and the neutralinos at the 
 electroweak scale. A large universal scalar mass would also imply that the sleptons
 also have few TeV size masses. Thus this set up would not lead to  stau \co which requires
that the lightest neutralino which we assume to be the LSP and the the lighter stau be in proximity
with a mass gap so that $(m_{\tilde \tau}- m_{\tilde \chi^0})/(m_{\tilde \tau}+m_{\tilde \chi^0})\leq 1/20$.
In high scale models stau \co can occur with  charginos and the neutralinos at the 
 electroweak scale if we lower the universal scalar mass so that the sleptons in general  have a mass
 comparable to the masses of the charginos and the neutralinos.  This leads us to non-universal SUGRA 
 models of a specific variety, i.e., where we consider non-universalities in the gaugino sector. 
 Specifically, if we consider the mass of the $SU(3)_C$ gaugino ($m_3$)  at the grand unification scale 
 to be much larger than the masses of the $U(1)_Y$ and $SU(2)_L$ gauginos ($m_1, m_2$), i.e., 
 $m_3>> m_1, m_2$.  In this class of models which are labeled $\tilde g$SUGRA~\cite{Akula:2013ioa},
 $m_0$ is chosen to be relatively low of size a few hundred GeV, while 
  $m_3$ is taken to be relatively large of size in the several  TeV.  The large $m_3$ mass  
  drives the squark masses to acquire TeV size masses through renormalization group evolution
  (for a review see~\cite{Ibanez:2007pf}), while
  the slepton masses remain largely unaffected~\cite{Akula:2013ioa}. This set up allows one to realize  stau \co 
  since both the neutralino and the stau lie in the sub TeV region and can lie close to each other. 

The  parameter space of this model is thus given by 
$m_0\,, A_0\,,m_1=m_2<<m_3\,$, $\tan\beta\,, \text{sign}(\mu)\,$,
where $A_0$ is the universal trilinear scalar coupling at the grand unification scale, $\tan\beta=\langle H_2\rangle/\langle H_1\rangle$, where $H_2$ gives mass to the up quarks and $H_1$ gives mass to the down quarks and the leptons, and sign$(\mu)$ is the sign of the Higgs mixing parameter which enters in the superpotential in the term $\mu H_1 H_2$.
For the multiparticle coannihilation parameter space, we relax the requirement that $m_1=m_2$, allowing $m_2$ to lie lower than $m_1$.  This brings the mass of the chargino and second neutralino closer to the stau and the LSP so that those particles also contribute to coannihilation. In this case we use the following parameter space for the model:
$m_0\,, A_0\,, m_2<m_1<<m_3\,$,  $\tan\beta\,, \text{sign}(\mu)\,$.
 Using the above input parameters, the sparticle spectrum is generated using \code{SoftSUSY 3.7.3}~\cite{Allanach:2001kg, Allanach:2016rxd} while the analysis of the relic density is done using \code{micrOMEGAs 4.3.1}~\cite{Belanger:2014vza}. SUSY Les Houches Accord formatted data files are processed using \code{PYSLHA}~\cite{Buckley:2013jua}.\

First we consider parameter regions of the $\tilde g$SUGRA model with the Higgs boson mass of $125\pm2$~GeV  where stau--LSP coannihilation gives rise to an LSP relic density within the known limit $\Omega h^2<0.128$. A sample set of such points is given in Tables~\ref{one_particle_inputs} and~\ref{one_particle_spectrum}, where Table~\ref{one_particle_inputs} gives the  input parameters and Table~\ref{one_particle_spectrum} gives the sparticle masses for those inputs.
As demonstrated in Table~\ref{one_particle_spectrum}, the parameter points in the stau  coannihilation region have a very small stau--neutralino mass gap $\Delta= (m_{\tilde \tau_1}- m_{\tilde \chi_1^0}) \sim20\GeV$. Such a small gap raises many challenges for discovery. In cases with such little energy available for decay jets, initial and final state radiation (ISR and FSR) events are often relied upon to produce a more detectible signal at colliders.
Next we consider a model of multipartcle \co among the neutralino (LSP), the stau (NLSP), and the chargino and second neutralino, which in this model remain nearly degenerate. The parameter points of Table~\ref{param_inputs} are chosen so as to satisfy the constraints on the Higgs boson mass, $m_h=125\pm 2$ GeV, and the relic density $\Omega h^2 <0.128$ and in such a way as to produce the mass hierarchy $m_{\chi^0_1}<m_{\tilde{\tau}}<m_{\chi^{\pm}_1}\sim m_{\chi_2^0}$.
Some of the sparticle masses corresponding to Table 3 are given in Table 4.

In Fig. 1 we exhibit the sparticle mass hierarchies generated by model point (a) of Table 1. Here the mass hierarchy 
of some of the low lying sparticles is:  $\tilde \chi_1^0 < \tilde \tau_1 < \tilde \chi_1^0 \simeq \tilde \chi_1^{\pm} 
< \tilde \nu_\tau < \tilde \nu_L < \tilde \ell_L <\tilde \tau_2$.  
In Fig. 2 we exhibit the sparticle mass hierarchies generated by model point (iii) of Table 3. Here the mass hierarchy 
of some of the low lying sparticles is similar to that for model point (a) of Table 1 except that  $\tilde \tau_1$,
$\tilde \chi_2^0$ and $\tilde \chi_1^{\pm}$ are very close in mass. As  discussed in~\cite{Feldman:2007zn}
the sparticle mass hierarchies including the mass gaps contain significant information regarding the 
nature of soft breaking at the grand unification  scale and Figs. 1 and 2 are an illustration of this phenomenon.
Thus observation of low lying sparticles and measurement of their masses will allow us to narrow down in 
a significant way the nature of the unified model from which the sparticle spectrum originates.

\section{Analysis for coannihilation models at  LHC at $\sqrt s =13$ TeV}

After a scan of the non-universal Supergravity (nuSUGRA) parameter space was performed to select benchmark points for each of the two coannihilation models satisfying the Higgs boson mass constraint, the relic density, and the desired neutralino, stau, and chargino mass hierarchies discussed in the previous section (Tables~\ref{one_particle_inputs} and~\ref{param_inputs}), those points are then used for a Monte Carlo analysis of LHC signal regions. This analysis was performed with the \code{MADGRAPH 2.4.2}~\cite{Alwall:2014hca} software system. First, the Feynman diagrams were calculated for all possible decays of the form $pp\to\text{SUSY SUSY}$, where ``$\text{SUSY}$" can be any MSSM particle. The analysis is configured to include both ISR and FSR jets. With the sparticle spectra of the benchmark points calculated by \code{SoftSUSY}, as well as the decay widths and branching ratios calculated by \code{SDECAY} and \code{HDECAY} operating within \code{SUSY-HIT}~\cite{Djouadi:2006bz}, \code{MADEVENT} was used to simulate 50,000 MSSM decay events for each benchmark point. Hadronization of resultant particles is handled by \code{PYTHIA 6.4.28}~\cite{Sjostrand:2006za}, and ATLAS detector simulation and event reconstruction is performed by \code{DELPHES 3.3.3}~\cite{Ovyn:2009tx}. A large set of search analyses were performed on the generated events for each benchmark point. The analyses used \code{ROOT 5.34.21}~\cite{Brun:1997pa} to implement the constraints of the
search region for the signal regions involving hadronic $\tau$ final states and other leptonic final states (see Section~\ref{sec:signal}).

To allow comparison to the background, all of the signal region analyses were applied to pre generated backgrounds published by the SNOWMASS group~\cite{Avetisyan:2013onh}. For each benchmark point, a calculated implied integrated luminosity allowed direct comparison to the backgrounds. Each individual background process from the SNOWMASS background set was scaled by its own implied integrated luminosity and combined to determine a total background count for each signal region. The various background samples are grouped according to the generated final state, with a collective notation given by
\begin{equation}
\begin{aligned}
J &= \{u,\bar{u},d,\bar{d},s,\bar{s},c,\bar{c},b,\bar{b}\}\,, \\
L &= \{e^{+},e^{-},\mu^{+},\mu^{-},\tau^{+},\tau^{-},\nu_e,\nu_{\mu},\nu_{\tau}\}\,, \\
B &= \{W^+,W^-,Z,\gamma,h^0\}\,,\\
T &= \{t,\bar{t}\}\,, \\
H &= \{h^0\}\,.
\label{bglabels}
\end{aligned}
\end{equation}
In general, events with gauge bosons and the SM Higgs boson in the final state are grouped into a single ``boson" (B) category. Thus, for example, the data set ``Bjj-vbf" represents production via vector boson fusion of a gauge boson or a Higgs boson with at least two additional light-quark jets.  The standard model background is displayed for two kinematic variables  $M_{\text{eff}}$(incl.) and $E_T^{\text{miss}}$ in Fig.~\ref{fig:bgs}.

\subsection{LHC production and signal definitions}  
The signal regions considered here comprise two major categories, based upon the sparticle whose decay signatures they are meant to capture.  The first category of signal regions includes signatures based on hadronically decaying taus, which are an expected result of stau decay.  The second category involves signatures of multiple light leptons, which are meant to search for the decays of charginos and heavy neutralinos.  Because both of the coannihilation regions under investigation have light staus and electroweakinos (i.e., charginos and neutralinos heavier than the lightest neutralino),
it is expected that both signal region categories are viable for the stau coannihilation models considered here.
The first signal region studied in this work involves at most 1 hadronically decaying tau in the final state. The selection criteria for one $\tau_h$ are based on an optimization of those defined in~\cite{Florez:2016lwi}. The second set of signal regions looks for at most 2 hadronically decaying taus in the final state. The selection criteria used are a modification of those in~\cite{Florez:2016lwi}, where the first (SC1) involves cuts on the transverse momenta of $\tau_h$ and the second (SC2) involves cuts on the effective mass, $m_{\text{eff}}$, defined as the sum of the missing transverse energy $E^{\text{miss}}_T$ and the transverse momenta of the two leading hadronic taus.  These signal regions are discussed in greater detail in Section \ref{tau_signals}.
Next we analyze electron and muon signal regions based on the work of~\cite{ATLAS:2016uwq}. One set of signal regions requires two leptons in the final state, comprising either a same flavor opposite sign pair, or a different flavor opposite sign pair, with increasing cuts on kinematic variables.  The second set requires three leptons in the final state, two of which form a same flavor opposite sign pair.  These are discussed further in section~\ref{lepton_signals}.
Using the techniques and signal regions described above, we analyze each of the benchmark points in Tables~\ref{one_particle_inputs} and~\ref{param_inputs} to identify a signal region with minimum required integrated luminosity for 5$\sigma$ S/$\sqrt{\text{B}}$ discovery of that point at the LHC.
   
\section{Signature analysis and results}\label{sec:signal}

In Tables~\ref{s_prod} and~\ref{m_prod} we give an analysis of the sparticle  production cross sections for the models under study. The cross section for all models is dominated by the production of the neutralino $\tilde\chi_2^0$ and chargino $\tilde\chi_1^\pm$. In nearly every model point the only decay mode of $\tilde\chi_2^0$ is via the channel $\tilde\chi_2^0\to\tilde\tau\tau$, while the primary decay of the chargino is via the channel $\tilde\chi_1^\pm\to\tilde\tau\nu_\tau$ (see Tables~\ref{sBR1} and~\ref{mBR1}). The stau always decays through one channel, $\tilde\tau\to\tilde\chi_1^0\tau$ (see Tables~\ref{sBR2} and~\ref{mBR2}), where the available phase space for the emitted tau is small, resulting in a soft tau production making it difficult to observe with low integrated luminosity.

\subsection{$\tau$-based signals}\label{tau_signals}

We start by discussing the 1$\tau$ signature search by applying the selection criteria given in~\cite{Florez:2016lwi}. It turns out that the calculated luminosity necessary for a 5$\sigma$ discovery  lies  beyond the maximum integrated luminosity achievable at the LHC. The main problem is the cut on the missing transverse energy. In~\cite{Florez:2016lwi}, 
a cut on $E_T^{\text{miss}}$ was made so that 
$E^{\text{miss}}_T>230$ GeV.  However, for the class of models we consider this cut is not optimal 
as illustrated in the left panel of Fig.~\ref{fig:figure9}. Here one finds that we begin to lose the signal for $E^{\text{miss}}_T>200$ GeV leading to a small signal to background ratio in this case. Further, the cut on 
$p_T(\tau_h)$ of ~\cite{Florez:2016lwi}, i.e., 
 $15<p_T(\tau_h)<35$ GeV applied on the hadronic tau transverse momentum is not optimal for the models considered here. The right panel of Fig.~\ref{fig:figure9} shows that signal is above the background in the range $20-90$ GeV. Thus increasing the range of cut on $p_T(\tau_h)$ will produce better results.     
The optimized cuts for the 1$\tau$ signature are displayed in Table~\ref{tab:SR1} including three variations: $1\tau$-A, $1\tau$-B and $1\tau$-C. They correspond to variations of the cut on $p_T(\tau_h)$.   
Table~\ref{tab:L1} gives the minimum integrated luminosity needed for a 5$\sigma$ discovery using these cuts on each of the benchmark points of Table~\ref{param_inputs} which correspond to the multipartcle coannihilation region. The best results are obtained for the cuts of $1\tau$-A where the luminosity ranges from 1510 to 2650 fb$^{-1}$, which is less than the optimal integrated luminosity achievable at the LHC, i.e., $\sim 3000$ fb$^{-1}$. Results obtained from $1\tau$-C show, for the most part, luminosities greater than $3000$ fb$^{-1}$ since the range of the cut on $p_T({\tau_h})$ extends to 150 GeV, which is above the value at which the signal generally begins dropping below background.   

Another $\tau$ signature of interest is that of two hadronically decaying taus in the final state. Here we adopt the signal regions of Table~\ref{tab:SR1} to the $2\tau$ case by considering two selection criteria SC1 and SC2 as shown in Table~\ref{tab:SR2}. The first selection criterion SC1 is a duplication of the cuts from Table~\ref{tab:SR1}, modified to require a second $\tau$, while 
in the selection criterion 
SC2 we introduce the variable $m_{\text{eff}}$, defined as the scalar sum of the missing transverse energy and the transverse momenta of the two leading hadronic taus, $m_{\text{eff}}=E_T^{\text{miss}}+p_T^{\tau_{1h}}+p_T^{\tau_{2h}}$. For completeness, we apply those cuts also to the 1$\tau$ signal regions and find that this improves our results from Table~\ref{tab:L1}.
Thus, the new set of $\tau$ based signal regions after inclusion of additional selection criteria SC1 and SC2 are presented in Table~\ref{tab:SR2}. Here we veto on electrons, muons, and b-jets. In this set, we have removed the cut on the pseudorapidity of the leading jet which was among the cuts for the 1$\tau$ signature in Table~\ref{tab:SR1}. Also, an upper bound has been placed on the $E^{\text{miss}}_T$ cut to suppress values where the signal drops below the background. In Fig.~\ref{fig:figure4} we exhibit the distributions in $p_T(j_1)$, the transverse momentum of the leading jet, and the effective mass $m_{\text{eff}}$ for the model point (xi). The signal appears to be above the background for lower $p_T(j_1)$ and $m_{\text{Eff}}$ values at which the cuts were applied (Table~\ref{tab:SR2}). 

In Table~\ref{tab:L3} we give the required minimum integrated luminosities for discovery for points (a)-(k) corresponding to the stau coannihilation model. The 1$\tau$ signature performs better than the 2$\tau$ for both SRs, SC1 and SC2, thus the reason this channel was omitted from table~\ref{tab:L3}. In 1$\tau$SC1 and 1$\tau$SC2 all points perform well except h and j. This is because both have the lowest production cross-section (see Table~\ref{s_prod}) and thus require higher integrated luminosities for discovery. Point (k) has the lowest integrated luminosity of 220 $\ifb$ in 1$\tau$SC1-C and point (d) has the highest  at 2960 $\ifb$ also in 1$\tau$SC1-A. 
Similarly, we computed the integrated luminosities for benchmark points (i)-(xi) corresponding to the multiparticle coannihilation model. Focusing on the 1$\tau$ signature, we notice an improvement compared to what was presented in Table~\ref{tab:L1}. Here all luminosities appear to be well below 3000 $\ifb$ in both SC1 and SC2 and are, thus, within reach of the HL-LHC. Despite having a poorer performance, the 2$\tau$ SR and in particular 2$\tau$SC2-A gives the lowest luminosity of 73 $\ifb$ for point (xi). It is worth noting that 70\% of the points (i)-(xi) have 1$\tau$SC1-C as the leading SR while 75\% of the points (a)-(k) that are listed have 1$\tau$SC1-C as the leading SR. 
It can be seen that introducing the kinematic variable $m_{\text{eff}}$ has improved our results for some regions while it didn't have much effect on others. For example, in the 2$\tau$SC2-A, B and C, where this variable is considered, an integrated luminosity less than 100 $\ifb$ is obtained for point (xi) which can be reached by the end of the current LHC run. In addition, for equivalent kinematic cuts, signal regions demanding a single hadronically decaying tau performed better than those demanding two taus for both the multiparticle coannihilation and stau coannihilation regions. It must be noted that the slight differences between the 1$\tau$ and 2$\tau$ cuts in SC2 is needed to give plausible results for the 2$\tau$ channel. Forcing exactly the same cuts produces overall unsatisfactory results for the 2$\tau$ channel. 
Analysis of the discovery potential for supersymmetry for the parameter space of Table~\ref{param_inputs}, using the selection criteria of Table~\ref{tab:SR2}, where the minimum integrated luminosity needed for $5\sigma$ discovery is displayed 
in Table 15.

The last SR we will investigate for the 2$\tau$ based signal is SR-2$\tau$SC3 given in Table~\ref{tab:SRSC3}. The reason for doing so is to try and accommodate most of the variables used by ATLAS in their searches. One of those variables is the quantity $m_{\text{T}\tau1}+m_{\text{T}\tau2}$ defined as the sum of the transverse masses of the leading and sub-leading taus, where $m_{\text{T}\tau}$ is calculated from the transverse momentum of the tau and $\mathbf{p}_{\rm T}^{\rm miss}$, so that
\begin{equation}
    m_{\rm T\tau}(\mathbf{p}_{\rm T\tau},\mathbf{p}_{\rm T}^{\rm miss})=
    \sqrt{2(p_{\rm T\tau}\,E_{\rm T}^{\rm miss}-\mathbf{p}_{\rm T\tau}\cdot\mathbf{p}_{\rm T}^{\rm miss})},
\label{MT}
\end{equation}
and $\Delta R(\tau_h,\tau_h)$ is the separation between the first two leading taus. Also here we keep our veto on the b-jets, electrons and muons. Table~\ref{tab:L2SC3} shows the integrated luminosities obtained for three variations of the transverse mass sum pertaining to the multi-particle coannihilation model points (i)-(xi). Signal region  2$\tau$SC3-A gives integrated luminosities as low as 670 $\ifb$ (for point xi). However, more than half of the listed points in table~\ref{tab:L2SC3} have 2$\tau$SC3-B as their leading SR. It is clear that for the most part, SC1 and SC2 give better results. \\
In Figs.~\ref{fig:figure4} to~\ref{fig:figure6} we exhibit the distributions in different kinematical variables for the multiparticle coannihilation model (xi) at 73 fb$^{-1}$ for signal region $2\tau$SC2-A and at 670 $\ifb$ for signal region 2$\tau$SC3-A in Fig.~\ref{fig:figure8}, where we plot the number of SUSY signal events (red) against the square root of the SM background (blue). The left panel of Fig.~\ref{fig:figure4} shows the distribution in the transverse momentum of the leading jet, $p_T(j_1)$ and the right panel shows the distribution in the effective mass $m_{\text{eff}}$. A similar analysis is done in Fig.~\ref{fig:figure5} for the transverse momentum of the leading hadronic tau, $p_T(\tau_{1h})$ (left panel), and  sub-leading $p_T(\tau_{2h})$ (right panel). In Fig.~\ref{fig:figure6} the same analysis is done but for the spatial separation between the subleading hadronic jet and the leading jet, $\Delta R(\tau_{2h}, j_1)$ in the left panel and the missing transverse energy $E^{\text{miss}}_T$ in the right panel. The histogram for the sum of the transverse masses of the first two leading hadronic tau jets is shown in the left panel of Fig.~\ref{fig:figure8} and the histogram for the spatial separation between the two leading hadronic taus $\Delta R(\tau_{h}, \tau_{h})$, which is effective in discriminating against back-to-back events such as multi-jet production or $Z$ decays, is exhibited in the left panel. The distributions for two kinematical variables are also plotted for point (iii) in the 2$\tau$ channel exhibited in Fig.~\ref{fig:figure7} and showing an excess of the signal over background events.

\subsection{$e$ and $\mu$-based signals}\label{lepton_signals}

In addition to the direct production of $\tau$ leptons due to the decay of stau particles, it is expected that decays of charginos and heavy neutralinos will result in detectible light leptons (electrons and muons) upon which further signal regions can be based.  To evaluate the effectiveness of these types of searches in regions of stau coannihilation and multiparticle coannihilation, benchmark models of  Tables 1 and 3 are evaluated against electroweakino
 signals designed to search for decays of $\widetilde\chi_1^+\widetilde\chi_1^-$ and $\widetilde\chi_1^\pm\widetilde\chi_2^0$~\cite{ATLAS:2016uwq}.  These signal regions are classified according to the number of signal leptons.
In the two lepton case, six signal regions are defined in two broad categories (see Table 18): signal regions labeled as 2l-SF require that the signal leptons form a same flavor, opposite sign (SFOS) pair, while signal regions labeled as 2l-DF require a different flavor, opposite sign (DFOS) pair.  The sub-categories  A, B, and C in Table 18
indicate different cuts on the kinematic variable $m_{\rm T2}$~\cite{Lester:1999tx, Barr:2003rg, Lester:2014yga}, which is defined as
\begin{equation}
    m_{\rm T2}=\min\left[\max\left(m_{\rm T}(\mathbf{p}_{\rm T}(\ell1),\mathbf{q}_{\rm T}),
    m_{\rm T}(\mathbf{p}_{\rm T}(\ell_2),\,\mathbf{p}_{\rm T}^{\text{miss}}-
    \mathbf{q}_{\rm T})\right)\right]\,
\end{equation}
where $\mathbf{q}_{\rm T}$ is an arbitrary vector chosen to find the appropriate minimum and $m_{\rm T}$ is the transverse mass given by 
\begin{equation}
    m_{\rm T}(\mathbf{p}_{\rm T1},\mathbf{p}_{\rm T2})=
    \sqrt{2(p_{\rm T1}\,p_{\rm T2}-\mathbf{p}_{\rm T1}\cdot\mathbf{p}_{\rm T2})}.
\end{equation}

In addition to cutting on $m_{\rm T2}$ and the missing transverse energy $E^{\text{miss}}_T$, the three lepton signal regions contain three jet vetoes, requiring that events contain no jets other than very soft jets in three jet categories: $b$-tagged jets ($b$-jet veto), jets which are not $b$-tagged and which have $|\eta|\leq2.4$ (light jet veto), and jets which are not $b$-tagged and which have $2.4\leq|\eta|\leq4.5$ (forward jet veto).  Finally, for the 2l-SF signal regions, there is a $Z$ veto which requires that the invariant mass of the SFOS lepton pair not lie within 10 GeV of the $Z$ mass.
For the three lepton case, two of the leptons are required to comprise a SFOS pair, with the third lepton allowed to have the same or different flavor. For the case where all three leptons are the same flavor, the SFOS pair is chosen to be that whose invariant mass is closest to the $Z$ mass. The three lepton case admits two signal regions A and B, with B representing tighter cuts on relevant kinematic variables (see Table 19).
Here, in addition to a veto on $b$-tagged jets, cuts are applied to the missing transverse energy, the transverse momentum of the third lepton, the transverse mass as defined above, and the invariant mass of the SFOS pair.

With these signal regions, it is possible to assess the discovery potential of stau coannihilation region parameter points based on the signal from electroweakino decays. Tables~\ref{tab:s_2l},~\ref{tab:m_2l}, and ~\ref{tab:2l_summary} below describe the results in terms of the integrated luminosity in $\ifb$ required for a $5\sigma$ discovery.  Results for the three lepton signal regions described in Table~\ref{tab:SR3l} are not displayed because it is found that in all cases, the required luminosity for discovery was much larger than for the two lepton cases, indeed larger than the 3000$\ifb$.  This is due to the fact that the decay events from these coannihilation regions almost never produce three final state leptons.
We find that for the leptonic signal regions, as mentioned earlier it is only the two lepton signals that give promising results.  Of these, the signal regions which require a SFOS pair perform much better than those requiring a DFOS pair.  Thus, this specific signal region topology is found to be the best leptonic signal for the stau and multiparticle coannihilation regions.  The remaining variation is upon kinematic cuts, in this case the cut on the variable $m_{\rm T2}$.  As expected for a kinematic cut, the softer cut of 2l-SF-A is optimal for lower mass benchmark points, while the harder cut 2l-SF-C is optimal for higher mass  points.  The intermediate signal region 2l-SF-B was not optimal for any case studied.
Figs.~\ref{MT2_1} and~\ref{MT2_2} display the $m_{\rm T2}$ and $E_T^{\text{miss}}$ kinematic variables for signal and combined background after cuts.  Fig.~\ref{MT2_1} gives counts in $m_{\rm T2}$ after the 2l-SF-A signal region cuts for models (k) and (c), models for which that signal region is optimal, displayed at the integrated luminosity calculated as necessary for discovery. Fig.~\ref{MT2_2} gives counts in the same signal region and for the same models, but this time in the $E_T^{\text{miss}}$ kinematic variable.

\subsection{Combined Signal Region Results}

As an overall view of the  signal regions considered and their success in discriminating between signal and background, we list in Tables~\ref{leadinga} and~\ref{leadingi} the leading and sub-leading signal regions  and the corresponding model points for the stau and multiparticle coannihilation regions, respectively.  Model points are listed in an 
ascending order of luminosity.
The analysis of Tables~\ref{leadinga} and~\ref{leadingi} shows that probing the supersymmetric signals originating
from the stau \co and multiparticle \co regions would be challenging. By the end of this year the CMS experiment
is expected to collect about 45-50 $\ifb$ of data~\cite{lum1} and one expects similar amount of data from the 
ATLAS experiment~\cite{lum2}.
One expects that by the time the LHC RUN II is over one may have a large enough data set to probe part of the
stau and multiparticle \co regions specifically model k of Table 23 which can be probed with 97 fb$^{-1}$ of integrated
luminosity and model xi of Table 24 which can be probed with 73 fb$^{-1}$ of integrated luminosity. 
Of course after the high luminosity LHC  upgrade, HL-LHC  is expected to 
collect up to 3 ab$^{-1}$ of data at a center-of-mass energy of 14 TeV. Thus with this data the full set of 
models  listed in Tables~23 and  24 
can be tested.  We note here that the dark matter constraints would become even more severe if 
the neutralino contributed only a fraction  of the dark matter density in the universe as is the case in multi-component
dark matter models (see e.g.,~\cite{Feldman:2010wy}).   One recent entry is the ultralight 
boson~\cite{Hui:2016ltb,Marsh:2015xka,Halverson:2017deq} needed to explain cosmology at small scales which 
could contribute part of the relic density of dark matter. In this case the mass gaps between the neutralino and the 
stau would have to be even narrower to reduce the dark matter relic density to a fraction of the observed one. 
One item not addressed in this analysis
and which needs further study, is the effect of pile up (for a review of these effects see~\cite{Aad:2015ina}). 
 Such an analysis is outside the scope of the current work
but would be a topic of further study.

\subsection{Stau \co  and direct detection of dark matter}
The analysis presented in Tables~\ref{leadinga} and~\ref{leadingi}  
 give us a set of models which are consistent with  the Higgs boson mass constraint and 
the constraints on the relic density consistent with the WMAP~\cite{Larson:2010gs}
 and the PLANCK  experiment~\cite{Ade:2015xua} 
 and arise 
from the stau or stau-chargino-second neutralino  \co regions.   
It is of interest to investigate if these models are discoverable in direct detection experiment.
For these models the neutralino is mostly a Bino. Thus the neutralino is a linear combination of four states
$\tilde \chi^0= \alpha \lambda^0 + \beta \lambda^3 + \gamma \tilde H_1+ \delta \tilde H_2$ where $\lambda^0, \lambda^3$
are the Bino, Wino  and $\tilde H_1, \tilde H_2$ are the Higgsinos.  For the models of Table 1,
$|\beta|\leq 0.003,  |\gamma|\leq 0.015, |\delta|\leq 0.002$.
while for models of Table 3, $|\beta|\leq 0.039,  |\gamma|\leq 0.014, |\delta|\leq 0.002$.
One finds that the Wino and the Higgsino content of the models of Tables 1 and 3 are small, and the neutralino
is essentially a Bino. This makes the neutralino-proton cross sections relatively small. 
 In Table~\ref{DM1} we present the spin independent and spin-dependent neutralino-proton cross sections 
 for these models. The analysis of Table~\ref{DM1} shows that the spin-independent neutralino-proton cross section
 though small and $O(10^{48}$cm$^{-2})$  still lies
  significantly above the neutrino floor~\cite{Strigari:2009bq} 
 which is the minimum threshold for detectability (see Fig. 12). Some of the models
 also lie close to the lower bounds that the future dark matter experiments 
 LUX-ZEPLIN~\cite{Schumann:2015wfa,Cushman:2013zza} 
 would be able to reach.

\section{Conclusions}
Supersymmetry is desirable for a number of theoretical as well as phenomenological reasons.
Supergravity unification provides a framework for high scale models with a small number of 
parameters in terms of which the properties of low energy effective theory can be computed. 
The observation of the Higgs boson mass at $\sim 125$ GeV implies that the loop correction to the tree
level Higgs boson mass  is large which in turn implies that the scale of weak scale supersymmetry 
lies in the TeV region. This makes the search for supersymmetry more challenging than initially 
thought. For high scale models, there is another aspect which makes the observation of 
supersymmetry challenging. This concerns dark matter. For high scale models one finds
that often the parameter space that gives the desired Higgs boson mass gives a neutralino
which is mostly a Bino. For a Bino type neutralino, one needs \co to achieve the appropriate
relic density consistent with the  WMAP and PLANCK experiment. This means that there must
be one or more sparticles close by to coannihilate with the neutralino. The relatively small 
mass gap between the neutralino and the coannihilating particles implies that the final states
in the  decay of the coannihilating particles must be soft and thus hard to detect.  
In this work we have addressed this question in the context of stau \co. We have analyzed two
types of models: one type which involves only a two particle \co between the neutralino and the stau,
and the second type where the neutralino coannihilates with a stau, a chargino and a second neutralino. 
We have carried out an extensive signature analysis including a variety of signatures including
one tau and two tau final states as well as $e$ and $\mu$ final states. Our analysis shows that 
a variety of signatures exist where the neutralino-stau \co and the neutralino-stau-chargino-second neutralino
coannihilation can be discovered with the total integrated luminosity expected at the LHC in future.
We also analyzed the 
spin -independent neutralino proton cross section. It is found that the cross section lies significantly 
above the neutralino floor and some parts of the parameter space may be accessible in future dark 
matter experiments such as LUX and ZEPLIN and XENON1T.

\textbf{Acknowledgments: }
Computational time allocation at the high-performance  Cluster353 at the Advanced Scientific Computing Initiative (ASCI) and the Discovery Cluster at Northeastern University is acknowledged. Also acknowledged are conversations with Bhaskar Dutta and with Toyoko Orimoto. 
This research was supported in part by the NSF Grants PHY-1314774 and PHY-1620575.

\section{Tables} 
  
\begin{table}[H]
\begin{center}
\begin{tabulary}{0.85\textwidth}{l|CCCCCC}
\hline\hline\rule{0pt}{3ex}
Model & $m_0$ & $A_0$ & $m_1=m_2$ & $m_3$ & $\tan\beta$ \\
\hline\rule{0pt}{3ex}
\!\!a. & 286 & -523 & 314 & 3015 & 10 \\
b. & 297 & -553 & 343 & 3246 & 10 \\
c. & 267 & -378 & 367 & 2911 & 10 \\
d. & 295 & -491 & 381 & 2821 & 13 \\
e. & 325 & -416 & 412 & 3156 & 14 \\
f. & 317 & -497 & 437 & 3065 & 14 \\
g. & 364 & -587 & 445 & 3728 & 14 \\
h. & 412 & -904 & 503 & 4688 & 13 \\
j. & 337 &  833 & 593 & 3626 & 15 \\
k. & 295 &  -551 & 302 & 3165 & 10 \\
\hline
\end{tabulary}\end{center}
\caption{Input parameters for representative stau coannihilation benchmark points.
All masses are in\GeV.
}
\label{one_particle_inputs}
\end{table}

\begin{table}[H]
\begin{center}
\begin{tabulary}{0.85\textwidth}{l|CCCCCCC}
\hline\hline\rule{0pt}{3ex}
Model  & $h^0$ & $\tilde\tau$ & $\tilde\chi_1^0$ & $\tilde\chi_1^\pm$ & $\tilde t$ & $\tilde g$ & $\Omega$ \\
\hline\rule{0pt}{3ex}
\!\!a. & 123.2 & 134.4 & 112.4 & 208.4 & 4522 & 6168 & 0.125 \\
 b. &  123.4 & 144.3 & 123.9 & 229.7 & 4842 & 6608 & 0.121 \\
 c. &  123.1 & 155.1 & 136.5 & 256.0 & 4376 & 5961 & 0.119 \\
 d. &  123.1 & 163.9 & 143.7 & 270.7 & 4244 & 5787 & 0.115 \\
 e. &  123.2 & 176.7 & 155.5 & 292.2 & 4720 & 6428 & 0.133 \\
 f. &  123.3 & 188.9 & 167.3 & 315.5 & 4584 & 6251 & 0.126 \\
 g. &  123.4 & 190.3 & 167.0 & 312.0 & 5506 & 7517 & 0.125 \\
 h. &  123.9 & 212.0 & 187.4 & 347.6 & 6775 & 9287 & 0.126 \\
 j. &  123.7 & 254.0 & 232.9 & 439.5 & 5422 & 7308 & 0.116 \\
 k. &  123.2 & 121.9 & 106.2 & 195.3 & 4732 & 6456 & 0.072 \\
\hline
\end{tabulary}\end{center}
\caption{The Higgs boson ($h^0$) mass, some relevant sparticle masses, and the relic density  for the stau coannihilation benchmark points of Table~\ref{one_particle_inputs}.  All masses are in $\GeV$.   }
\label{one_particle_spectrum}
\end{table}

\begin{table}[H]
\begin{center}
\begin{tabulary}{0.85\textwidth}{l|CCCCCC}
\hline\hline\rule{0pt}{3ex}
Model & $m_0$ & $A_0$ & $m_1$ & $m_2$ & $m_3$ & $\tan\beta$ \\
\hline\rule{0pt}{3ex}
i.   & 345  & 68  & 394  & 287 & 3690  & 10 \\
ii.  & 385  & 152 & 403 & 290 & 3972 & 12 \\
iii. & 318  & 248 & 357 & 249 & 2973 & 12 \\
iv.  & 386  & -47  & 401  & 284 & 3809  & 13 \\
v.   & 367  & 78  & 409  & 290 & 3550  & 13 \\
vi.  & 423  & -19 & 431 & 314 & 4396 & 13 \\
vii. & 353 & 202 & 427 & 298 & 3351 & 13 \\
viii.& 390  & -161   & 440 & 308 & 3864 & 13 \\
ix.  & 321 & 246  & 423 & 296 & 3328 & 10 \\
x.   & 432 & 264 & 494 & 350 & 4234 & 15 \\
xi.  & 304 & -745 & 260 & 221 & 2793 & 11 \\
\hline
\end{tabulary}\end{center}
\caption{Input parameters for representative stau--chargino coannihilation benchmark points. All masses are in $\GeV$.}
\label{param_inputs}
\end{table}
%

\begin{table}[H]
\begin{center}
\begin{tabulary}{0.85\textwidth}{l|CCCCCCC}
\hline\hline\rule{0pt}{3ex}
Model & $h^0$ & $\tilde\tau$ & $\tilde\chi_1^0$ & $\tilde\chi_1^\pm$ & $\tilde t$ & $\tilde g$ & $\Omega$ \\
\hline\rule{0pt}{3ex}
i.   &  123.8 & 161.4 & 142.3 & 171.7 & 5511 & 7468 & 0.124 \\
ii.  &  123.8 & 166.5 & 144.6 & 169.7 & 5912 & 8007 & 0.127 \\
iii. &  123.2 & 150.2 & 130.4 & 151.4 & 4521 & 6098 & 0.114 \\
iv.  & 123.6 & 166.9 & 145.0 & 167.9 & 5677 & 7698 & 0.115 \\
v.   &  123.6 & 171.4 & 150.2 & 177.8 & 5320 & 7201 & 0.120 \\
vi.  &  123.8 & 176.5 & 154.7 & 183.9 & 6488 & 8808 & 0.107 \\
vii. &  123.5 & 179.6 & 159.3 & 188.2 & 5045 & 6818 & 0.117 \\
viii. &  123.8 & 182.8 & 162.2 & 188.5 & 5742 & 7797 & 0.103  \\
ix. & 123.6 & 175.1& 157.3 & 185.9 & 5011 & 6773 & 0.121 \\
x. &  123.5 & 206.5 & 184.0 & 219.0 & 6272 & 8492 & 0.101 \\
xi.&  123.1 & 121.6 & 89.9 & 131.5 & 4212 & 5756 & 0.125 \\
\hline
\end{tabulary}\end{center}
\caption{
The Higgs boson mass, some relevant sparticle masses, and the relic density  for the stau--chargino coannihlation benchmark points of Table~\ref{param_inputs}.  All masses are in $\GeV$.}
\label{param_spectrum}
\end{table}

\begin{table}[H]
\begin{center}
\begin{tabulary}{0.85\textwidth}{l|CCCCC}
\hline\hline\rule{0pt}{3ex}
Model & full SUSY & $\tilde\chi_2^0\tilde\chi_1^\pm$ & $\tilde\chi_1^+\tilde\chi_1^-$ & $\tilde\tau^+\tilde\tau^-$ & $\tilde\tau\tilde\nu_\tau$\\
\hline\rule{0pt}{3ex}
\!\!a.& 2.09 & 2.57 & 0.62 & 0.04 & 0.03 \\
b. & 1.48 & 0.88 & 0.43 & 0.03 & 0.03 \\
c. & 1.01 & 0.58 & 0.29 & 0.02 & 0.02 \\
d. & 0.79 & 0.47 & 0.23 & 0.02 & 0.01 \\
e. & 0.59 & 0.35 & 0.17 & 0.01 & 0.01 \\
f. & 0.44 & 0.26 & 0.13 & 0.01 & 0.007 \\
g. & 0.46 & 0.28 & 0.13 & 0.01 & 0.008 \\
h. & 0.31 & 0.18 & 0.09 & 0.007 & 0.006 \\
j. & 0.12 & 0.07 & 0.03 & 0.003 & 0.002 \\
k. & 2.65 & 1.61 & 0.79 & 0.06 & 0.04 \\
\hline
\end{tabulary}\end{center}
\caption{SUSY production cross sections in picobarns  for stau coannihilation benchmark points of Table~\ref{one_particle_inputs}.}
\label{s_prod}
\end{table}  

\begin{table}[H]
\begin{center}
\begin{tabulary}{0.85\textwidth}{l|CCCCC}
\hline\hline\rule{0pt}{3ex}
Model & full SUSY & $\tilde\chi_2^0\tilde\chi_1^\pm$ & $\tilde\chi_1^+\tilde\chi_1^-$ & $\tilde\tau^+\tilde\tau^-$ & $\tilde\tau\tilde\nu_\tau$\\
\hline\rule{0pt}{3ex}
i.    & 3.99 & 2.57 & 1.26 & 0.02 & 0.03\\
ii.   & 4.12 & 2.68 & 1.32 & 0.02 & 0.02\\
iii.  & 6.17 & 4.02 & 1.98 & 0.03 & 0.03\\
iv.   & 4.25 & 2.78 & 1.37 & 0.02 & 0.02\\
v.    & 3.48 & 2.27 & 1.11 & 0.02 & 0.02\\
vi.   & 3.08 & 2.01 & 0.98 & 0.02 & 0.02\\
vii.  & 2.84 & 1.84 & 0.90 & 0.02 & 0.01\\
viii. & 2.81 & 1.83 & 0.90 & 0.01 & 0.01\\
ix.   & 3.01 & 1.93 & 0.94 & 0.02 & 0.02\\
x.    & 1.63 & 1.06 & 0.52 & 0.01 & 0.01\\
xi.   & 10.19 & 6.65 & 3.29 & 0.06 & 0.04\\
\hline
\end{tabulary}\end{center}
\caption{SUSY production cross sections in picobarns  for multiparticle coannihilation benchmark points of Table~\ref{param_inputs}.}
\label{m_prod}
\end{table}

\begin{table}[H]
\begin{center}
\begin{tabulary}{0.85\textwidth}{l|CC}
\hline\hline\rule{0pt}{3ex}
Model & $\tilde\chi_2^0\to\tilde\tau\tau$ & $\tilde\chi_1^\pm\to\tilde\tau\nu_\tau$\\
\hline\rule{0pt}{3ex}
\!\!a.& 1.00 &  1.00\\
b. & 1.00 &  1.00\\
c. & 0.99 &  0.99\\
d. & 1.00 &  1.00\\
e. & 1.00 &  1.00\\
f. & 1.00 &  1.00\\
g. & 1.00 &  1.00\\
h. & 0.99 &  0.99 \\
j. & 0.75 &  0.75 \\
k. & 1.00 & 0.99  \\
\hline
\end{tabulary}\end{center}
\caption{Branching ratios for dominant decays of $\tilde\chi_1^\pm$ and $\tilde\chi_2^0$ for stau coannihilation benchmark points of Table~\ref{one_particle_inputs}.}
\label{sBR1}
\end{table}

\begin{table}[H]
\begin{center}
\begin{tabulary}{0.85\textwidth}{l|CCC}
\hline\hline\rule{0pt}{3ex}
Model & $\tilde\chi_2^0\to\tilde\tau\tau$ & $\tilde\chi_2^0\to\tilde\chi_1^0\tau^{+}\tau^{-}$ & $\tilde\chi_1^\pm\to\tilde\tau\nu_\tau$\\
\hline\rule{0pt}{3ex}
i.  & 1.00 & 0.00 & 1.00 \\
ii. & 1.00 & 0.00 & 1.00 \\
iii. & 0.00 & 0.99 & 1.00 \\
iv.  & 0.00 & 0.99  & 1.00  \\
v.   & 1.00 & 0.00 & 1.00 \\
vi.  & 1.00 & 0.00 & 1.00 \\
vii. & 1.00 & 0.00 & 1.00 \\
viii. & 1.00 & 0.00  & 1.00 \\
ix.& 1.00 & 0.00 & 1.00 \\
x. & 1.00 & 0.00 & 1.00 \\
xi. & 1.00 & 0.00 & 1.00 \\
\hline
\end{tabulary}\end{center}
\caption{Branching ratios for dominant decays of $\tilde\chi_1^\pm$ and $\tilde\chi_2^0$ for multiparticle coannihilation benchmark points of Table~\ref{param_inputs}.}
\label{mBR1}
\end{table}

\begin{table}[H]
\begin{center}
\begin{tabulary}{0.85\textwidth}{l|CCCCC}
\hline\hline\rule{0pt}{3ex}
Model & $\tilde\tau\to\tilde\chi_1^0\tau$ & $\tilde\nu_\tau\to\tilde\chi_1^0\nu_\tau$ & $\tilde\nu_\tau\to\tilde\chi_2^0\nu_\tau$ & $\tilde\nu_\tau\to\tilde\chi_1^{\pm}\tau$ & $\tilde\nu_\tau\to\tilde\tau W^{\pm}$ \\
\hline\rule{0pt}{3ex}
\!\!a.& 1.00 & 0.38 & 0.07 & 0.13 & 0.42 \\
b. & 1.00 & 0.42 & 0.02 & 0.03 & 0.53 \\
c. & 1.00 & 0.47 & 0.00 & 0.00 & 0.53 \\
d. & 1.00 & 0.17 & 0.01 & 0.03 & 0.79 \\
e. & 1.00 & 0.12 & 0.01 & 0.02 & 0.85 \\
f. & 1.00 & 0.13 & 0.002 & 0.004 & 0.87 \\
g. & 1.00 & 0.11 & 0.006 & 0.13 & 0.87 \\
h. & 1.00 & 0.12 & 0.00 & 0.00 & 0.88 \\
j. & 1.00 & 0.10 & 0.00 & 0.00 & 0.90 \\
k. & 1.00 & 0.31 & 0.08 & 0.15 & 0.46 \\
\hline
\end{tabulary}\end{center}
\caption{Branching ratios for dominant decays of $\tilde\tau$ and $\tilde\nu_\tau$ for stau coannihilation benchmark points of Table~\ref{one_particle_inputs}.}
\label{sBR2}
\end{table}

\begin{table}[H]
\begin{center}
\begin{tabulary}{0.85\textwidth}{l|CCCCC}
\hline\hline\rule{0pt}{3ex}
Model & $\tilde\tau\to\tilde\chi_1^0\tau$ & $\tilde\nu_\tau\to\tilde\chi_1^0\nu_\tau$ & $\tilde\nu_\tau\to\tilde\chi_2^0\nu_\tau$ & $\tilde\nu_\tau\to\tilde\chi_1^{\pm}\tau$ & $\tilde\nu_\tau\to\tilde\tau W^{\pm}$ \\
\hline\rule{0pt}{3ex}
i.   & 1.00 & 0.16 & 0.27 & 0.56 & 0.00\\
ii.  & 1.00 & 0.11 & 0.23 & 0.46 & 0.20 \\
iii. & 1.00 & 0.12 & 0.25 & 0.51 & 0.12\\
iv.  & 1.00 & 0.09 & 0.20 & 0.41 & 0.30 \\
v.   & 1.00 & 0.10 & 0.21 & 0.42 & 0.28 \\
vi.  & 1.00 & 0.09 & 0.18 & 0.37 & 0.36\\
vii. & 1.00 & 0.11 & 0.22 & 0.44 & 0.23 \\
viii.& 1.00 & 0.09 & 0.20 & 0.41 & 0.29 \\
ix.  & 1.00 & 0.18 & 0.27 & 0.55 & 0.00 \\
x.   & 1.00 & 0.07 & 0.15 & 0.30 & 0.48\\
xi.  & 1.00 & 0.12 & 0.23 & 0.47 & 0.18\\
\hline
\end{tabulary}\end{center}
\caption{Branching ratios for dominant decays of $\tilde\tau$ and $\tilde\nu_\tau$ for multiparticle coannihilation benchmark points of Table~\ref{param_inputs}.}
\label{mBR2}
\end{table}

\begin{table}[H]
	\begin{center}
	\begin{tabulary}{\linewidth}{l|c|c|c}
    \hline\hline
	Requirement & \multicolumn{3}{c}{Value} \\
    \hline
    & $1\tau$-A & $1\tau$-B & $1\tau$-C \\
    \cline{2-4}
    $E^{\text{miss}}_T\text{ (GeV)}>$
    & 130 & 130 & 130 \\
    $p_T(j_1)\text{ (GeV)}>$
    & 100 & 100 & 100 \\
    $|\eta(j_1)|<$
    & 2.5 & 2.5 & 2.5 \\
    $p_T(\tau_h)\text{ (GeV)}>$
    & 15 & 15 & 20 \\
    $p_T(\tau_h)\text{ (GeV)}<$
    & 50 & 80 & 150 \\
    $|\eta(\tau_h)|<$
    & 2.3 & 2.3 & 2.3 \\
    $\Delta R(\tau_h,j_1)>$
    & 0.4 & 0.4 & 0.4 \\
    \hline
	\end{tabulary}\end{center}
	\caption{The selection criteria used for the signal regions with one hadronically decaying tau in the final state and a veto on electrons, muons and b-jets.  The angles are in rad.}
	\label{tab:SR1}
\end{table}
\begin{table}[H]
	\centering
	\begin{tabulary}{0.5\linewidth}{l|CCC}
    \hline\hline
	Model & {$\mathcal{L}$ for $5\sigma$ discovery in $1\tau$-A} & $\mathcal{L}$ for $5\sigma$ discovery in $1\tau$-B & $\mathcal{L}$ for $5\sigma$ discovery in $1\tau$-C\\
	\hline
    i.    & 1510 & 1810 & 2520\\
    ii.   & 1550 & 1800 & 2630\\
    iii.  & 1550 & 1910 & 2730\\
    iv.   & 1580 & 2020 & 2930\\
    v.    & 1800 & 2260 & \dd\\
    vi.   & 2010 & 2290 & \dd\\
    vii.  & 2010 & 2330 & \dd\\
    viii. & 2090 & 2340 & \dd\\
    ix.   & 2400 & 2880 & \dd\\
    x.    & 2650 & \dd & \dd\\
    xi.   & 1610 & 1420 & 1720\\
	\hline
	\end{tabulary}
	\caption{Analysis of the discovery potential for supersymmetry for the parameter space of Table~\ref{param_inputs}, using the selection criteria of Table~\ref{tab:SR1}, where the minimum integrated luminosity needed for $5\sigma$ discovery is given in$\ifb$. Here and in the tables following two dots (\dd) indicate that the 
	 minimum integrated luminosity needed for $5\sigma$ discovery exceeds $3000\ifb$.	}
	\label{tab:L1}
\end{table}

\begin{table}[H]
	\begin{center}
	\begin{tabulary}{0.85\linewidth}{l|c|c|c|c|c|c}
    \hline\hline
	Requirement & \multicolumn{6}{c}{SC1} \\
    \hline
    & $1\tau$-A & $1\tau$-B & $1\tau$-C & $2\tau$-A & $2\tau$-B & $2\tau$-C\\
    \cline{2-7}
    $p_T(j_1)\text{ (GeV)}>$
    & 20 & 20 & 20 & 20 & 20 & 20 \\
    $p_T(j_1)\text{ (GeV)}<$
    & 100 & 100 & 100 & 100 & 100 & 100 \\
    $p_T(\tau_{1h})\text{ (GeV)}>$
    & 20 & 20 & 20 & 20 & 20 & 20 \\
    $p_T(\tau_{1h})\text{ (GeV)}<$
    & 50 & 70 & 90 & 50 & 70 & 90 \\
    $p_T(\tau_{2h})\text{ (GeV)}>$
    & - & - & - & 20 & 20 & 20 \\
    $p_T(\tau_{2h})\text{ (GeV)}<$
    & - & - & - & 40 & 50 & 60  \\
    $|\eta(\tau_{1h})|<$
    & 1.2 & 1.2 & 1.2 & 1.2 & 1.2 & 1.2  \\
    $|\eta(\tau_{2h})|<$
    & - & - & - & 1.0 & 1.0 & 1.0 \\
    $\Delta R(\tau_{1h},j_1)>$
    & 0.6 & 0.6 & 0.6 & 0.6 & 0.6 & 0.6  \\
    $\Delta R(\tau_{1h},j_1)<$
    & 1.8 & 1.8 & 1.8 & 1.8 & 1.8 & 1.8 \\
    $\Delta R(\tau_{2h},j_1)>$
    & - & - & - & 2.3 & 2.3 & 2.3  \\
    $\Delta R(\tau_{2h},j_1)<$
    & - & - & - & 3.3 & 3.3 & 3.3\\
    $N(\tau_h)$
    & 1 & 1 & 1 & 2 & 2 & 2  \\
    \hline
Requirement &  \multicolumn{6}{c}{SC2} \\
    \hline
    & $1\tau$-A & $1\tau$-B & $1\tau$-C & $2\tau$-A & $2\tau$-B & $2\tau$-C \\
    \cline{2-7}
    $p_T(j_1)\text{ (GeV)}>$
    & 20 & 20 & 20 & - & - & -\\
    $p_T(j_1)\text{ (GeV)}<$
    & 200 & 200 & 200 & 110 & 110 & 110 \\
    $|\eta(\tau_{1h})|<$
    &  1.2 & 1.2 & 1.2 & 1.4 & 1.4 & 1.4 \\
    $|\eta(\tau_{2h})|<$
    &  - & - & - & 1.0 & 1.0 & 1.0 \\
    $\Delta R(\tau_{1h},j_1)>$
    &  0.6 & 0.6 & 0.6 & 0.8 & 0.8 & 0.8 \\
    $\Delta R(\tau_{1h},j_1)<$
    &  1.8 & 1.8 & 1.8 & 1.8 & 1.8 & 1.8 \\
    $\Delta R(\tau_{2h},j_1)>$
    & - & - & - & 2.3 & 2.3 & 2.3 \\
    $\Delta R(\tau_{2h},j_1)<$
    &  - & - & - & 3.3 & 3.3 & 3.3 \\
    $m_{\text{eff}}>$
    & 120 & 130 & 140 & 110 & 110 & 110 \\
    $m_{\text{eff}}<$
    &  200 & 250 & 300 & 250 & 350 & 450 \\
    $N(\tau_h)$
    &  1 & 1 & 1 & 2 & 2 & 2 \\
    \hline
    \end{tabulary}\end{center}
	\caption{The selection criteria (SC) used for the signal regions with the 1$\tau$ and 2$\tau$ signatures. The SRs SC1 and SC2 have a common cut on the missing transverse energy of 100 GeV $< E^{\text{miss}}_T <$ 200 GeV, with a veto on electrons, muons and b-jets. The dashes mean that the kinematical variable is not applicable to the corresponding SR.  The angles are in rad.	
	}
	\label{tab:SR2}
\end{table}

\begin{table}[H]
	\centering
	\begin{tabulary}{\linewidth}{l|ccc|ccc}
    \hline\hline
	& \multicolumn{3}{c|}{$\mathcal{L}$ for $5\sigma$ discovery in SC1}
    & \multicolumn{3}{c}{$\mathcal{L}$ for $5\sigma$ discovery in SC2}  \\
	\hline
	Model & $1\tau$-A & $1\tau$-B & $1\tau$-C & $1\tau$-A & $1\tau$-B & $1\tau$-C \\
	\hline
    a.  & 786 & 487 & 303  &  745 & 383 & 313  \\
    b.  & 1310 & 674 & 416 &  1120 & 621 & 536  \\
    c.  & 2760 & 1280 & 756 &  2460 & 1340 & 1020  \\
    d.  & 2960 & 1490 & 967  &  2840 & 1470 & 1050\\
    e. & \dd & 2860 & 1700&  \dd & 2170 & 1660  \\
    f. & \dd & \dd & 2210&  \dd & \dd & 2340 \\
    g. & \dd & \dd & 2460 &  \dd & \dd & 2340\\
    k. & 427 & 279 & 220 & 644 & 349 & 299 \\
	\hline
	\end{tabulary}
	\caption{Analysis of the discovery potential for supersymmetry for the parameter space of Table~\ref{one_particle_inputs}, using the selection criteria of Table~\ref{tab:SR2}, where the minimum integrated luminosity needed for $5\sigma$ discovery is given in $\ifb$.  Points $h$ and $j$ are not listed because the integrated
	luminosity for discovery exceeds 3000 $\ifb$.	
	Only 1$\tau$ signal regions are displayed, as those are the signal regions which give luminosities for discovery in the reasonable range.}
	\label{tab:L3}
\end{table}

\begin{table}[H]
	\centering
	\begin{tabulary}{1.1\linewidth}{l|CCC|CCC|CCC|CCC}
    \hline\hline
	& \multicolumn{6}{c|}{$\mathcal{L}$ for $5\sigma$ discovery} & \multicolumn{6}{c}{$\mathcal{L}$ for $5\sigma$ discovery} \\
	& \multicolumn{6}{c|}{in SC1} & \multicolumn{6}{c}{in SC2} \\
	\hline
	Model & $1\tau$-A & $1\tau$-B & $1\tau$-C & $2\tau$-A & $2\tau$-B & $2\tau$-C & $1\tau$-A & $1\tau$-B & $1\tau$-C & $2\tau$-A & $2\tau$-B & $2\tau$-C\\
	\hline
    i.    & 1020 & 704 & 625 & 1090 & 2460 & 1290 & 1040 & 715 & 694 & 477 & 579 & 582 \\
    ii.   & 501 & 380 & 292 & - & - & - & 536 & 370 & 352 & - & - & - \\
    iii.  & 637 & 512 & 472 & 458 & 1030 & 542 & 827 & 669 & 648 & 200 & 243 & 244 \\
    iv.   & 677 & 575 & 532 & - & - & - & 941 & 695 & 666 & \dd & \dd & \dd \\
    v.    & 654 & 475 & 411 & 1440 & 1440 & 1710 & 1070 & 894 & 898 & 631 & 765 & 770 \\
    vi.   & 898 & 853 & 650 & - & - & - & 1170 & 743 & 693 & 1810 & 2200 & 2210 \\
    vii.  & 730 & 605 & 508 & \dd & \dd & \dd & 1190 & 825 & 870 & \dd & \dd & \dd \\   
    viii. & 1040 & 746 & 660 & 2200 & \dd & 2060 & 1250 & 889 & 842 & 2170 & 1170 & 1170 \\
    ix. & 1190 & 713 & 661 & \dd & \dd& \dd & 1610 & 1020 & 1020 & 842 & 575 & 578 \\
    x.  & 1430 & 1230 & 1090 & \dd & \dd & \dd & 1950 & 1340 & 1270 & 2880 & \dd & \dd \\
    xi.  & 265 & 169 & 144 & 168 & 378 & 199 & 176 & 143 & 119 & 73 & 89 & 90 \\
	\hline
	\end{tabulary}
	\caption{Analysis of the discovery potential for supersymmetry for the parameter space of Table~\ref{param_inputs}, using the selection criteria of Table~\ref{tab:SR2}, where the minimum integrated luminosity needed for $5\sigma$ discovery is given in$\ifb$. The dashes mean that zero events have passed the applied cuts.}
	\label{tab:L2}
\end{table}

\begin{table}[H]
	\begin{center}
	\begin{tabulary}{\linewidth}{l|c|c|c}
    \hline\hline
	Requirement & \multicolumn{3}{c}{SR-SC3} \\
    \hline
    & $2\tau$-A & $2\tau$-B & $2\tau$-C \\
    \cline{2-4}
    $E^{\text{miss}}_T\text{ (GeV)}>$
    & 100 & 100 & 100 \\
    $E^{\text{miss}}_T\text{ (GeV)}<$
    & 200 & 200 & 200 \\
    $p_T(j_1)\text{ (GeV)}<$
    & 180 & 180 & 180 \\
    $m_{\text{eff}}\text{ (GeV)}>$
    & 130 & 130 & 130 \\
    $m_{\text{eff}}\text{ (GeV)}<$
    & 200 & 200 & 200 \\
    $m_{\text{T}\tau1}+m_{\text{T}\tau2}>$
    & 100 & 100 & 50 \\
    $m_{\text{T}\tau1}+m_{\text{T}\tau2}<$
    & 200 & 300 & 500 \\
    $\Delta R(\tau_h,\tau_h)>$
    & 2.5 & 2.5 & 2.5 \\
    $\Delta R(\tau_h,\tau_h)<$
    & 3.5 & 3.5 & 3.5 \\
    \hline
	\end{tabulary}\end{center}
	\caption{The selection criteria used for the signal regions SR-SC3 with 2 hadronically decaying taus in the final state ($N(\tau_h)=2$) and a veto on electrons, muons and b-jets.}
	\label{tab:SRSC3}
\end{table}

\begin{table}[H]
	\centering
	\begin{tabulary}{0.6\linewidth}{l|CCC}
    \hline\hline
	Model & $\mathcal{L}$ for $5\sigma$ discovery in $2\tau$-SC3-A & $\mathcal{L}$ for $5\sigma$ discovery in $2\tau$-SC3-B & $\mathcal{L}$ for $5\sigma$ discovery in $2\tau$-SC3-C\\
	\hline
    i.    & 1240 & 1090 & 1430\\
    ii.   & 1820 & 1560 & 1720\\
    iii.  & 1170 & 1180 & 1730\\
    v.    & 1640 & 1450 & 2130\\
    vii.  & 1710 & 1540 & 2270\\
    viii. & 2870 & 2510 & \dd\\
    x.    & 2690 & 2700 & \dd\\
    xi.   & 670 & 674 & 991 \\
    \hline
	\end{tabulary}
	\caption{Analysis of the discovery potential for supersymmetry for the parameter space of Table~\ref{param_inputs}, using the selection criteria of Table~\ref{tab:SRSC3}, where the minimum integrated luminosity needed for $5\sigma$ discovery is given in$\ifb$.  Models $iv, vi,$ and $ix$ are not listed because
	the minimum integrated luminosity needed for $5\sigma$ discovery exceeded $3000\ifb$.}
	\label{tab:L2SC3}
\end{table}

\begin{table}[H]
	\begin{center}
	\begin{tabulary}{\linewidth}{l|c|c|c|c|c|c}
    \hline\hline
	Requirement & \multicolumn{3}{c|}{SF} & \multicolumn{3}{c}{DF} \\
    \hline
    & $2l$-SF-A & $2l$-SF-B & $2l$-SF-C & $2l$-DF-A & $2l$-DF-B & $2l$-DF-C \\
    \cline{2-7}
    $E^{\text{miss}}_T\text{ (GeV)}>$
    & 100 & 100 & 100 & 100 & 100 & 100 \\
    light jet $p_T\text{ (GeV)}<$
    & 20 & 20 & 20 & 30 & 30 & 30 \\
    $b$-jet $p_T\text{ (GeV)}<$
    & 20 & 20 & 20 & 20 & 20 & 20 \\
    forward jet $p_T\text{ (GeV)}<$
    & 30 & 30 & 30 & 30 & 30 & 30 \\
    $|m_{\ell\ell}-m_Z|\text{ (GeV)}>$
    & 10 & 10 & 10 & - & - & - \\
    $m_{T2}\text{ (GeV)}>$
    & 90 & 120 & 150 & 90 & 120 & 150 \\
	\hline
	\end{tabulary}\end{center}
	\caption{The selection criteria used for the signal regions related to the 2 lepton signature, based upon the 2 lepton signal regions from~\cite{ATLAS:2016uwq}. Here and in the Tables following SF stands for 
	same flavor opposite sign lepton pair  and DF stands for different flavor opposite sign lepton pair.
	A dash denotes a cut which is not applicable to the given signal region.}
	\label{tab:SR2l}
\end{table}

\begin{table}[H]
	\begin{center}
	\begin{tabulary}{\linewidth}{l|c|c}
    \hline\hline
	Requirement & \multicolumn{2}{c}{Value} \\
    \hline
    & $3l$-A & $3l$-B \\
    $E^{\text{miss}}_T\text{ (GeV)}>$
    & 120 & 100 \\
    $p_T(\ell_3)\text{ (GeV)}>$
    & 30 & 80 \\
    $m_T\text{ (GeV)}<$
    & 110 & 110 \\
    $m_{\text{SFOS}}\text{ (GeV)}$
    & $\notin[21.2, 101.2]$ & $>101.2$ \\
    $N(\text{b-jet})$
    & 0 & 0 \\
	\hline
	\end{tabulary}\end{center}
	\caption{The selection criteria used for the signal regions related to the 3 lepton signature, based upon the 2 lepton signal regions from~\cite{ATLAS:2016uwq}.}
	\label{tab:SR3l}
\end{table}

\begin{table}[H]
	\centering
	\begin{tabulary}{0.8\linewidth}{l|CCC}
    \hline\hline
	& \multicolumn{3}{c}{$\mathcal{L}$ for $5\sigma$ discovery in 2l-SF} \\
	\hline
Model & $2l$-SF-A & $2l$-SF-B & $2l$-SF-C \\
	\hline
a.& 187 & 266 & 266 \\
b.& 362 & 420 & 441 \\
c.& 165 & 188 & 169 \\
d.& 781 & 953 & 884 \\
e.& 1480 & 1630 & 1700 \\
f.& 1110 & 1380 & 1250 \\
g.& 1850 & 1850 & 1790 \\
h.& 1860 & 2050 & 1660 \\
j.& 2160 & 2250 & 1880 \\
k.& 97 & 185 & 225 \\
	\hline
	\end{tabulary}
	\caption{Analysis of the discovery potential for supersymmetry for the parameter space of Table~\ref{one_particle_inputs}, using the 2 lepton same flavor (SF) selection criteria of Table~\ref{tab:SR2l}, where the minimum integrated luminosity needed for $5\sigma$ discovery is given in$\ifb$. The different flavor (DF) signal regions are omitted due to poor performance (i.e. requiring over $3000\ifb$ of integrated luminosity for discovery).}
	\label{tab:s_2l}
\end{table}

\begin{table}[H]
	\centering
	\begin{tabulary}{0.8\linewidth}{l|CCC}
    \hline\hline
	& \multicolumn{3}{c}{$\mathcal{L}$ for $5\sigma$ discovery in 2l-SF} \\
	\hline
Model & $2l$-SF-A & $2l$-SF-B & $2l$-SF-C \\
	\hline
i.    & 545 & 623 & 696 \\
ii.   & 315 & 306 & 273 \\
iii.  & 181 & 271 & 238 \\
iv.   & 640 & 843 & 934 \\
v.    & 1410 & 1460 & 1690 \\
vi.   & 1090 & 1610 & 1500 \\
vii.  & 944 & 1450 & 1510 \\
viii. & 732 & 1090 & 1190 \\
ix.   & 360 & 487 & 624 \\
xi.    & 224 & 450 & 547 \\
	\hline
	\end{tabulary}
	\caption{Analysis of the discovery potential for supersymmetry for the parameter space of Table~\ref{param_inputs}, using the 2 lepton same flavor (SF) selection criteria of Table~\ref{tab:SR2l}, where the minimum integrated luminosity needed for $5\sigma$ discovery is given in$\ifb$. The different flavor (DF) signal regions are omitted due to poor performance (i.e. requiring over $3000\ifb$ of integrated luminosity for discovery).
	Model x is not listed because
	the minimum integrated luminosity needed for $5\sigma$ discovery exceeded $3000\ifb$	
	}
	\label{tab:m_2l}
\end{table}

\begin{table}[H]
\begin{center}
\begin{tabulary}{0.85\textwidth}{rCC|rCC}
\hline\hline\rule{0pt}{3ex}
Model & Leading SR & $\mathcal{L}$ (fb$^{-1}$) & Model & Leading SR & $\mathcal{L}$ (fb$^{-1}$)\\
\hline\rule{0pt}{3ex}
a. & 2l-SF-A & 187  & i. & 2l-SF-A & 454\\
b. & 2l-SF-A & 362  & ii. & 2l-SF-C & 273\\
c. & 2l-SF-A & 165  & iii. & 2l-SF-A & 181\\
d. & 2l-SF-A & 781  & iv. & 2l-SF-A & 640\\
e. & 2l-SF-A & 1480 & v. & 2l-SF-A & 1410\\
f. & 2l-SF-A & 1110 & vi. & 2l-SF-A & 1090\\
g. & 2l-SF-C & 1790 & vii. & 2l-SF-A & 944\\
h. & 2l-SF-C & 1660 & viii.& 2l-SF-A & 732 \\
j. & 2l-SF-C & 1880 & ix. & 2l-SF-A & 360\\
k. & 2l-SF-A & 97   & x. & 2l-SF-C & \dd\\
   &         &      & xi. & 2l-SF-A & 224\\
\hline
\end{tabulary}\end{center}
\caption{Integrated luminosity for SUSY production in the leading and sub-leading leptonic ($e$ and $\mu$) signal regions of Tables~\ref{tab:SR2l} and~\ref{tab:SR3l} for the benchmark points of Tables~\ref{one_particle_inputs} and~\ref{param_inputs}.}
\label{leading}
\label{tab:2l_summary}
\end{table}

\begin{table}[H]
\begin{center}
\begin{tabulary}{0.85\textwidth}{lCCCCC}
\hline\hline\rule{0pt}{3ex}
Model & Leading SR & $\mathcal{L}$ (fb$^{-1}$) & Sub-leading SR & $\mathcal{L}$ (fb$^{-1}$)\\
\hline\rule{0pt}{3ex}
\!\!k. & 2l-SF-A & 97 & 2l-SF-B & 185 \\
c. & 2l-SF-A & 165 & 2l-SF-C & 169 \\
a.& 2l-SF-A & 187 & 2l-SF-B & 266 \\
b.& 2l-SF-A & 362 & $1\tau$-SC2-C & 416\\
d.& 2l-SF-A & 781 & 2l-SF-C & 884\\
f.& 2l-SF-A & 1110 & 2l-SF-C & 1250\\
e.& 2l-SF-A & 1480 & 2l-SF-B & 1630\\
g.& 2l-SF-C & 1790 & 2l-SF-A & 1850\\
h.& 2l-SF-C & 1660 & 2l-SF-A & 1860\\
j.& 2l-SF-C & 1880 & 2l-SF-A & 2160\\
\hline
\end{tabulary}\end{center}
\caption{
The overall minimum integrated luminosities needed for $5\sigma$ discovery using the
leading and the sub-leading signal regions for stau coannihilation models of Table~\ref{one_particle_inputs}, including the $\tau$ based signal regions discussed in Section~\ref{tau_signals} as well as the $e$ and $\mu$ based signal regions discussed in Section~\ref{lepton_signals}.}
\label{leadinga}
\end{table}

\begin{table}[H]
\begin{center}
\begin{tabulary}{0.85\textwidth}{lCCCCC}
\hline\hline\rule{0pt}{3ex}
Model & Leading SR & $\mathcal{L}$ (fb$^{-1}$) &  Sub-leading SR & $\mathcal{L}$ (fb$^{-1}$)\\
\hline\rule{0pt}{3ex}
\!\!xi. & 2$\tau$-SC2-A & 73 & 2$\tau$-SC2-B & 89 \\
iii. & 2l-SF-A & 181 & 2$\tau$-SC2-A & 200 \\
ii. & 2l-SF-C & 273 & 2l-SF-B & 306 \\
ix. & 2l-SF-A & 360 & 2l-SF-B & 487 \\
v. & 1$\tau$-SC1-C & 411 & 1$\tau$-SC1-B & 475 \\
i. & 2$\tau$-SC2-A & 477 & 2l-SF-A & 545 \\
vii. & 1$\tau$-SC1-C & 508 & 1$\tau$-SC1-B & 605 \\
iv. & 1$\tau$-SC1-C & 532 & 1$\tau$-SC1-B & 575 \\
vi. & 1$\tau$-SC1-C & 650 & 1$\tau$-SC2-C & 693 \\
viii. & 1$\tau$-SC1-C & 660 & 2l-SF-A & 732 \\
x. & 1$\tau$-SC1-C & 1090 & 1$\tau$-SC1-B & 1230 \\
\hline
\end{tabulary}\end{center}
\caption{The overall minimum integrated luminosities needed for $5\sigma$ discovery using the 
leading and sub-leading signal regions for stau coannihilation models of Table~\ref{param_inputs}, including the $\tau$ based signal regions discussed in Section~\ref{tau_signals} as well as the $e$ and $\mu$ based signal regions discussed in Section~\ref{lepton_signals}.}
\label{leadingi}
\end{table}

\begin{table}[H]
\begin{center}
\begin{tabulary}{0.85\textwidth}{lCC}
\hline\hline\rule{0pt}{3ex}
Model & $\sigma^{\text{SI}}_{p,\chi^0_1}\times 10^{48}$ & $\sigma^{\text{SD}}_{p,\chi^0_1}\times 10^{46}$ \\
\hline\rule{0pt}{3ex}
\!\!a.& 0.92 & 4.77 \\
b.& 0.80 & 3.67 \\
c.& 1.08 & 5.60 \\
d.& 0.80 & 6.35 \\
e.& 0.58 & 4.43 \\
f.& 0.64 & 4.82 \\
g.& 0.40 & 2.39 \\ 
h.& 0.27 & 1.03 \\
j.& 0.53 & 3.52 \\ 
k.& 1.22 & 0.25 \\ 
  &      &      \\
\hline
\end{tabulary}
\begin{tabulary}{0.85\textwidth}{lCC}
\hline\hline\rule{0pt}{3ex}
Model & $\sigma^{\text{SI}}_{p,\chi^0_1}\times 10^{48}$ & $\sigma^{\text{SD}}_{p,\chi^0_1}\times 10^{45}$ \\
\hline\rule{0pt}{3ex}
i. & 1.33 & 3.02 \\
ii.& 1.97 & 3.54 \\
iii.& 0.94 & 1.03 \\
iv. & 1.94 & 2.87 \\
v.& 1.74 & 2.45 \\
vi.& 2.90 & 5.66 \\
vii.& 1.46 & 1.93 \\
viii.& 2.02 & 3.28 \\
ix.& 0.96 & 1.88 \\
x.& 3.01 & 4.77 \\
xi.& 1.11 & 1.53 \\
\hline
\end{tabulary}
\caption{Three left columns: Proton--neutralino spin-independent ($\sigma^{\text{SI}}_{p,\chi^0_1}$) and spin-dependent ($\sigma^{\text{SD}}_{p,\chi^0_1}$) cross-sections in units of cm$^{-2}$ for the 10 benchmark points of Table
~\ref{one_particle_inputs}. Three right columns:
Proton--neutralino spin-independent ($\sigma^{\text{SI}}_{p,\chi^0_1}$) and spin-dependent ($\sigma^{\text{SD}}_{p,\chi^0_1}$) cross-sections in units of cm$^{-2}$ for the 11 benchmark points of Table~\ref{param_inputs}.}
\label{DM1}
\end{center}
\end{table}

\section{Figures}

\begin{figure}[H]
\begin{center}
	\includegraphics[width=0.45\textwidth]{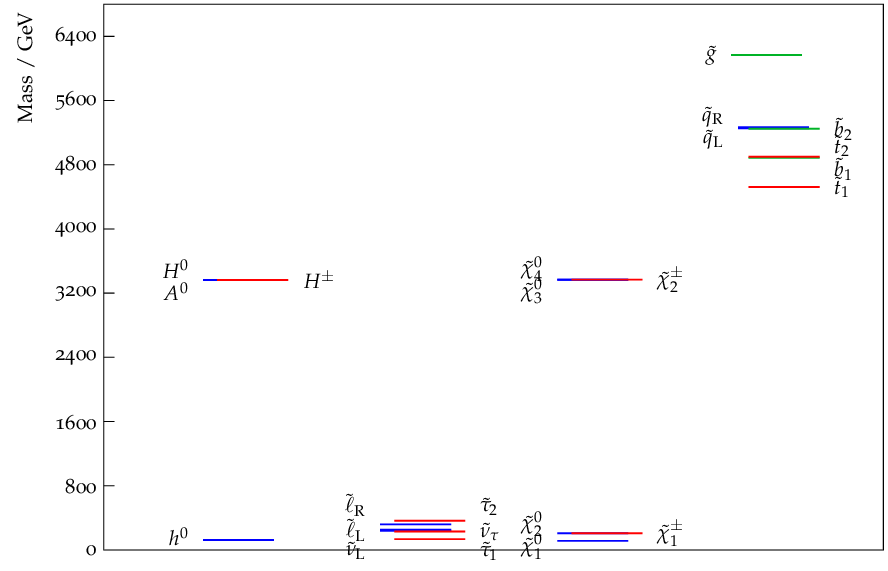}
	\includegraphics[width=0.45\textwidth]{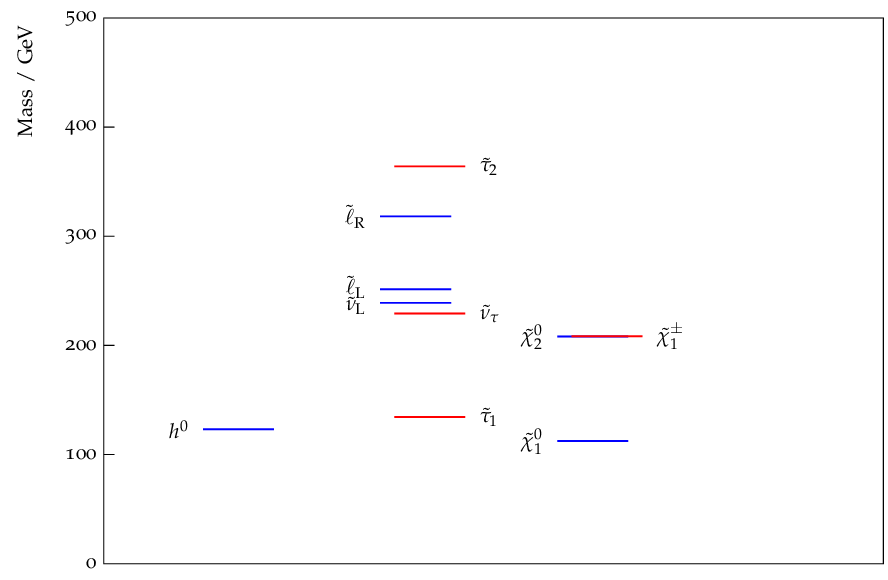}
\end{center}
    \caption{An exhibition of the sparticle mass hierarchy for stau coannihilation model (a). Left panel: Full spectrum.  Right panel: Only sparticles with mass $<500$ GeV.}
    \label{hierarchy4}
\end{figure}

\begin{figure}[H]
\begin{center}
	\includegraphics[width=0.45\textwidth]{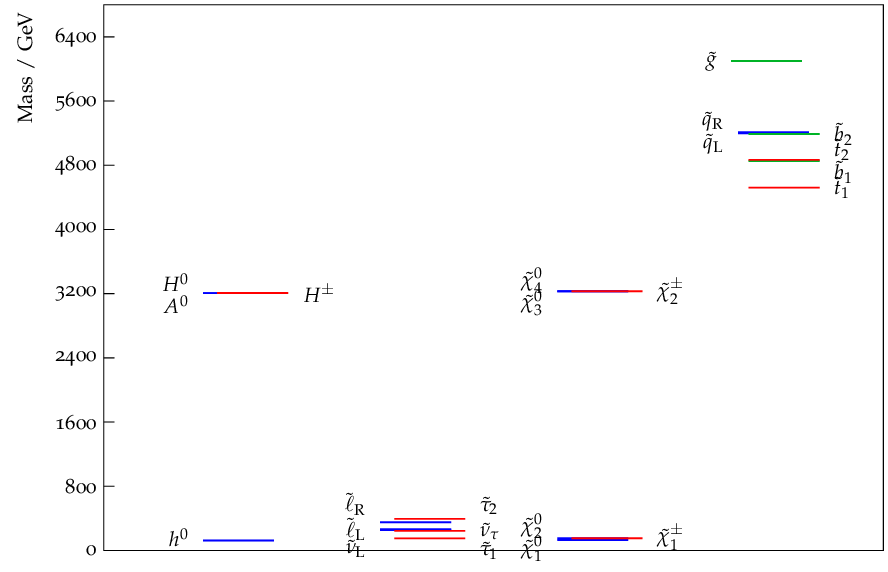}
		\includegraphics[width=0.45\textwidth]{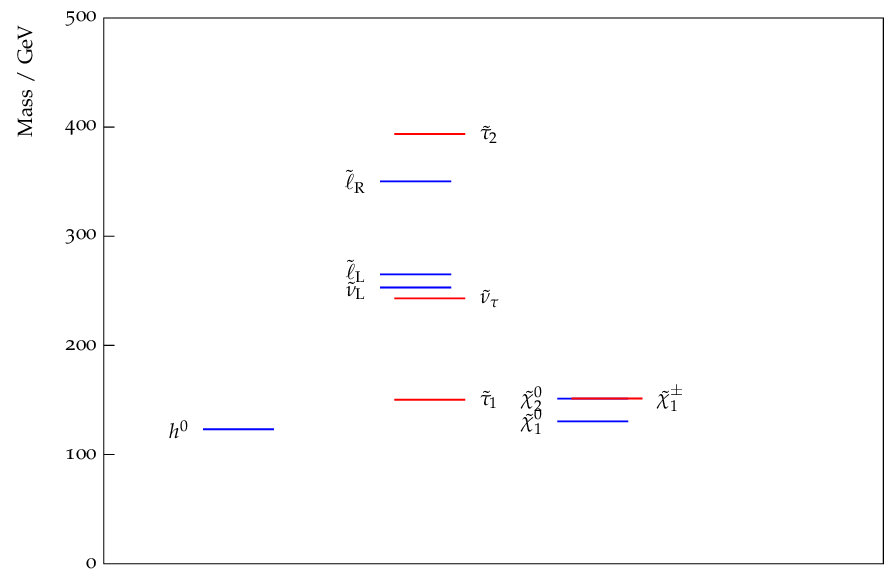}
\end{center}
    \caption{An exhibition of the sparticle mass hierarchy for multiparticle coannihilation model (iii). Left panel: Full spectrum.  Right panel: Only sparticles with mass $<500$ GeV.}
    \label{hierarchy1}
\end{figure}

\begin{figure}[h]
    \centering
	\includegraphics[width=0.45\textwidth]{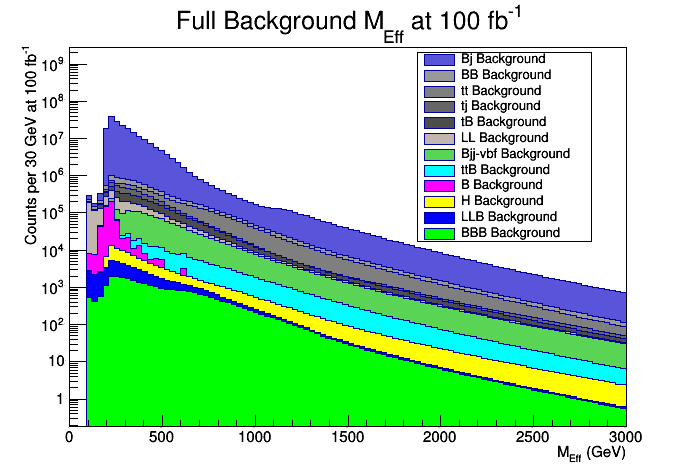}
    \includegraphics[width=0.45\textwidth]{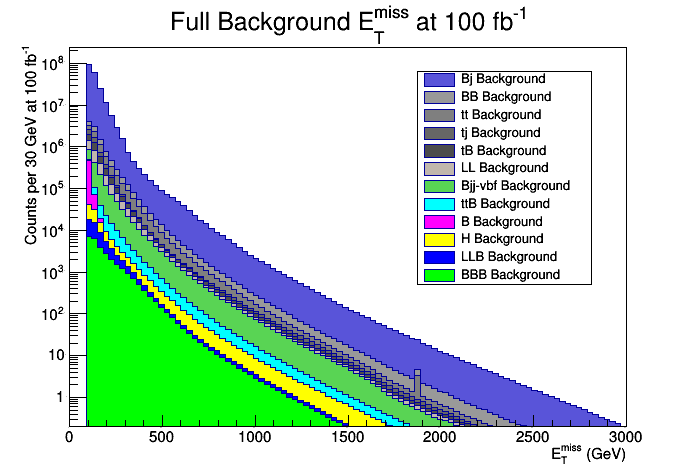}
    \caption{Full SNOWMASS standard model background~\cite{Avetisyan:2013onh} after triggering cuts and a cut of $E_T^{\text{miss}}\geq100\GeV$, broken into final states and scaled to $100\ifb$. The top panel gives $M_{\text{eff}}$(incl.) and the bottom panel gives $E_T^{\text{miss}}$. Individual data sets are labeled according to Eq. \ref{bglabels}.}
	\label{fig:bgs}
\end{figure}

\begin{figure}[htp]
\centering
\includegraphics[width=0.4\textwidth]{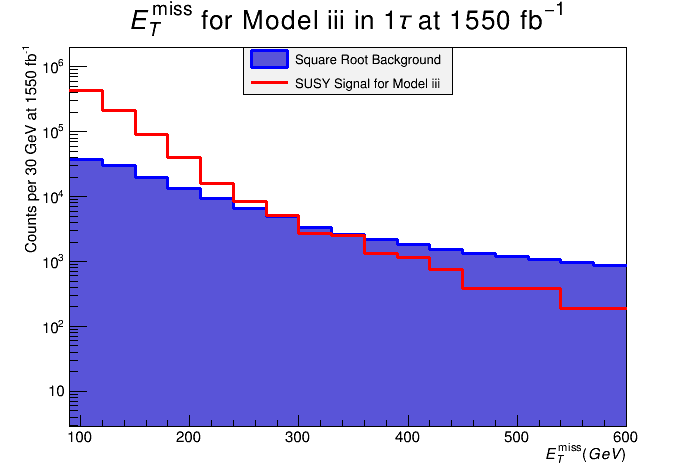}
\includegraphics[width=0.4\textwidth]{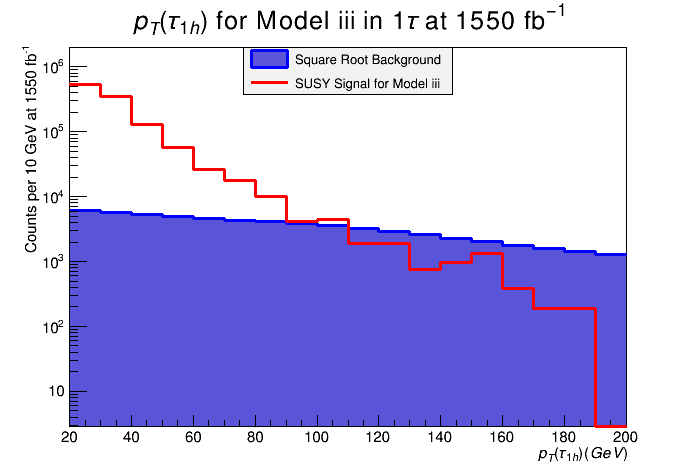}
\caption{Left panel: Distribution in $E^{\text{miss}}_T$ for the $1\tau$ signal region for multiparticle coannihilation model (iii) prior to any cuts. Plotted is the number of
counts for the SUSY signal per 30 GeV and the square root of the total SM SNOWMASS background. The analysis is done at 1550 fb$^{-1}$ of integrated luminosity, which gives a 5$\sigma$ discovery in this signal region. Right panel: The same analysis as in the left panel but for $p_T(\tau_h)$ with counts per 10 GeV.}
\label{fig:figure9}
\end{figure}

\begin{figure}[htp]
\centering
\includegraphics[width=.4\textwidth]{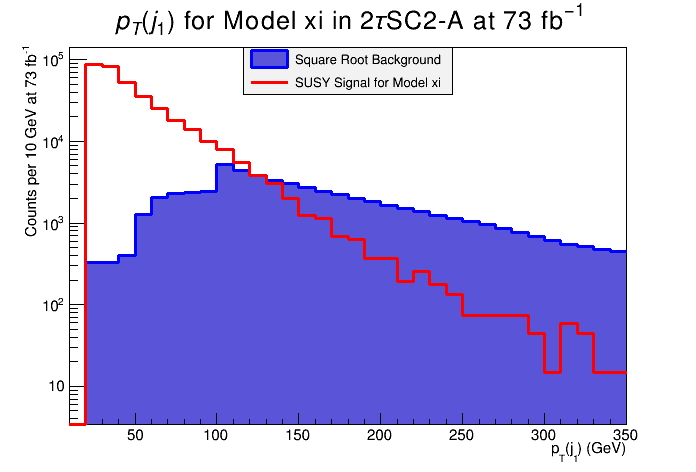}
\includegraphics[width=.4\textwidth]{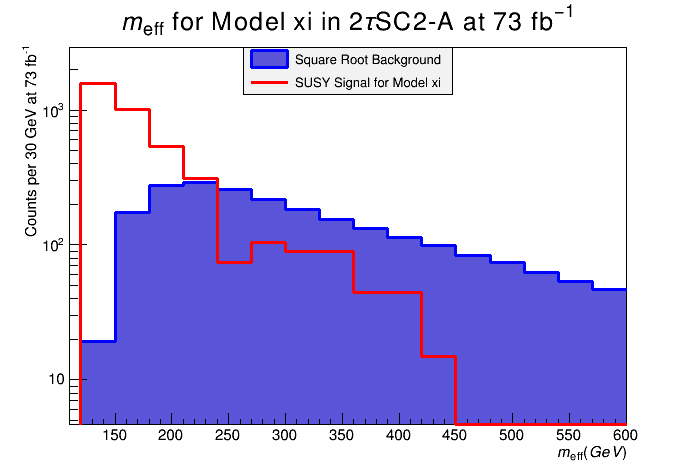}
\caption{Left panel: Distribution in $p_T(j_1)$ for the $2\tau$SC2-A signal region for multiparticle coannihilation model (xi) prior to any cuts. Plotted is the number of counts for the SUSY signal per 10 GeV and the square root of the total SM SNOWMASS backgrounds. The analysis is done at 73 fb$^{-1}$ of integrated luminosity, which gives a 5$\sigma$ discovery in this signal region. Right panel: The same analysis as in the left panel but for $m_{\text{eff}}$ with counts for the SUSY signal per 30 GeV.}
\label{fig:figure4}
\end{figure}

\begin{figure}[htp]
\centering
\includegraphics[width=.4\textwidth]{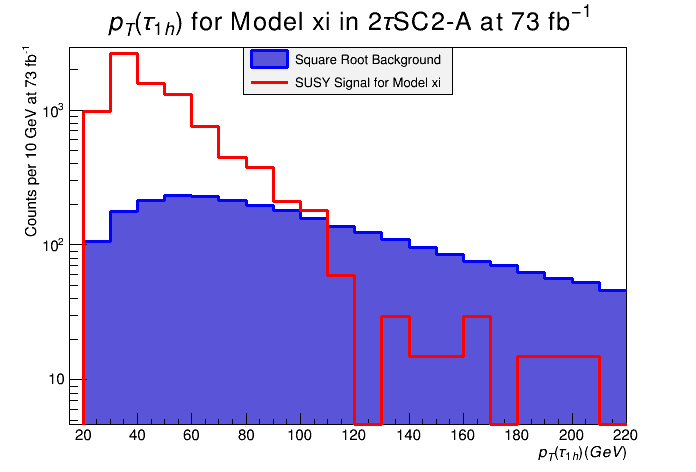}
\includegraphics[width=.4\textwidth]{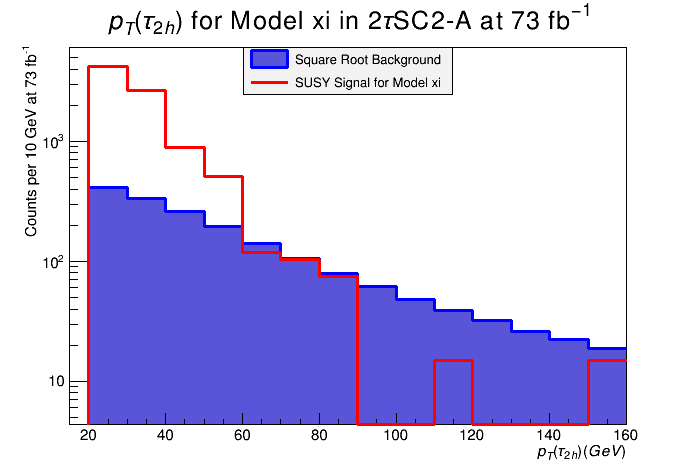}
\caption{Left panel: Distribution in $p_T(\tau_{1h})$ for the $2\tau$SC2-A signal region for the multiparticle coannihilation model (xi) prior to any cuts. Plotted is the number of counts for the SUSY signal per 10 GeV and the square root of the total SM SNOWMASS backgrounds. The analysis is done at 73 fb$^{-1}$ of integrated luminosity, which gives a 5$\sigma$ discovery in this signal region. Right panel: The same analysis as in the left panel but for $p_T(\tau_{2h})$.}
\label{fig:figure5}
\end{figure}

\begin{figure}[htp]
\centering
\includegraphics[width=.4\textwidth]{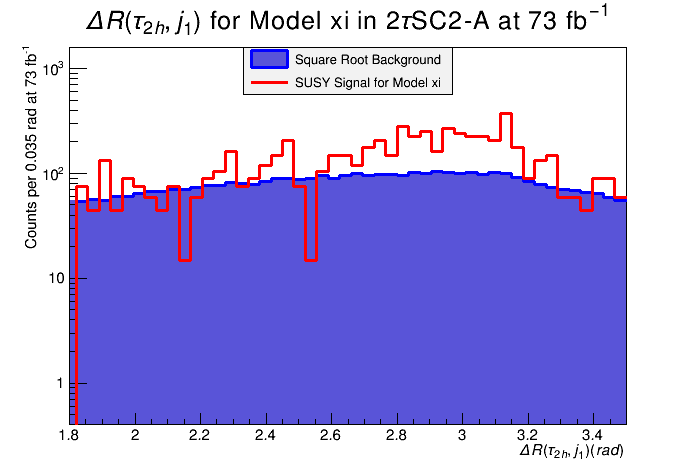}
\includegraphics[width=.4\textwidth]{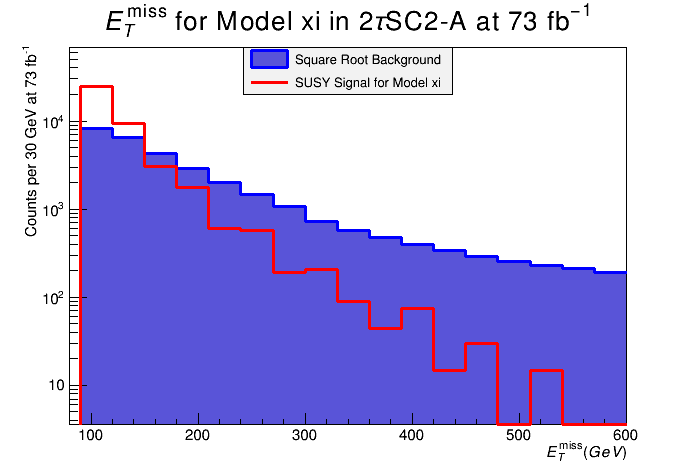}
\caption{Left panel: Distribution in $\Delta R(\tau_{2h}, j_{1})$ for the $2\tau$SC2-A signal region for the multiparticle coannihilation model (xi) prior to any cuts. Plotted is the number of counts for the SUSY signal per 0.035 rad and the square root of the total SM SNOWMASS backgrounds. The analysis is done at 73 fb$^{-1}$ of integrated luminosity, which gives a 5$\sigma$ discovery in this signal region. Right panel: The same analysis as in the left panel but for $E^{\text{miss}}_T$ with counts for the SUSY signal per 30 GeV.}
\label{fig:figure6}
\end{figure}

\begin{figure}[htp]
\centering
\includegraphics[width=.4\textwidth]{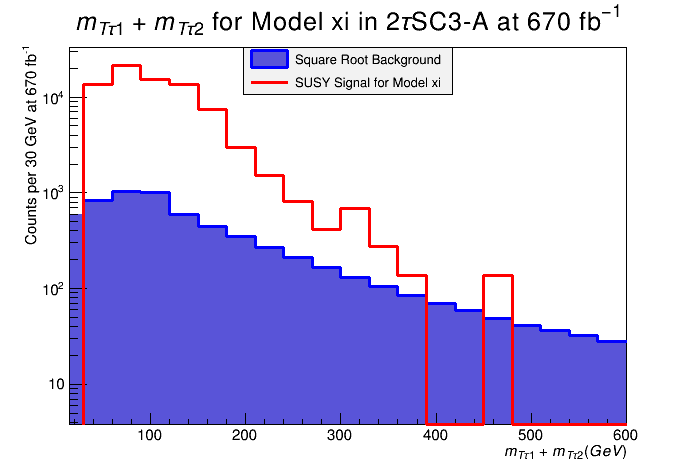}
\includegraphics[width=.4\textwidth]{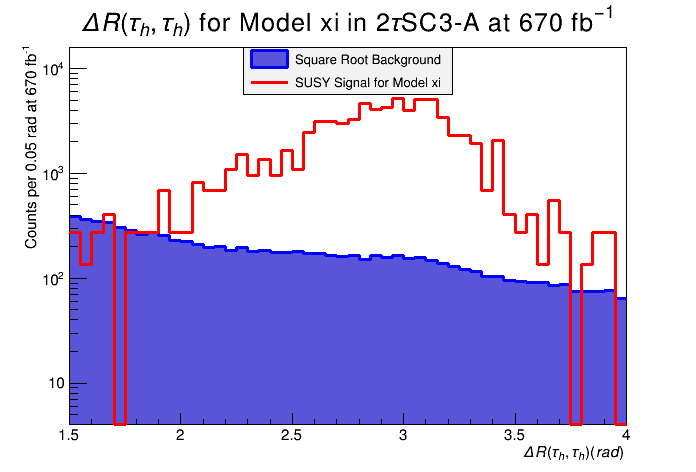}
\caption{Left panel: Distribution in $m_{T\tau1}+m_{T\tau2}$ for the $2\tau$SC3-A signal region for the multiparticle coannihilation model (xi) prior to any cuts. Plotted is the number of counts for the SUSY signal per 30 GeV and the square root of the total SM SNOWMASS backgrounds. The analysis is done at 670 fb$^{-1}$ of integrated luminosity, which gives a 5$\sigma$ discovery in this signal region. Right panel: The same analysis as in the left panel but for $\Delta R (\tau_h,\tau_h)$ with counts for the SUSY signal per 0.05 rad.}
\label{fig:figure8}
\end{figure}

\begin{figure}[htp]
\centering
\includegraphics[width=.4\textwidth]{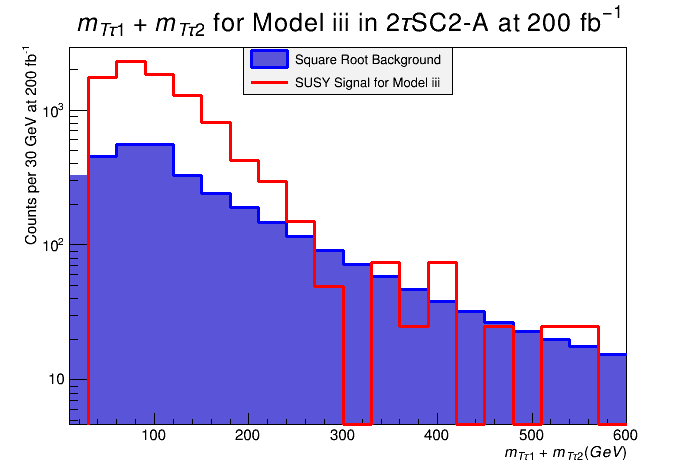}
\includegraphics[width=.4\textwidth]{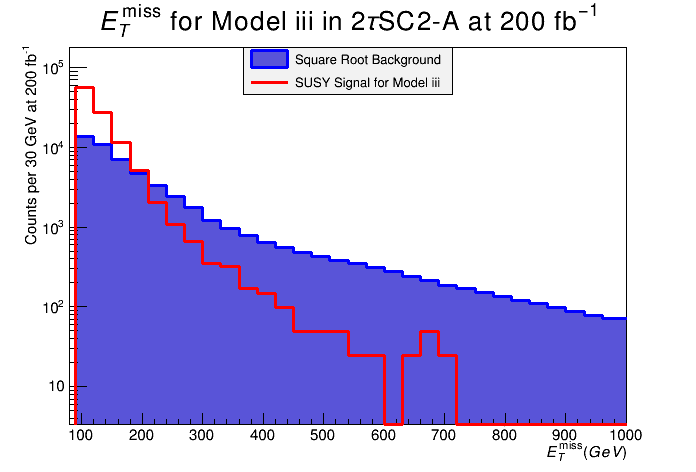}
\caption{Left panel: Distribution in $m_{T\tau1}+m_{T\tau2}$ for the $2\tau$SC2-A signal region for the multiparticle coannihilation model (iii) prior to any cuts. Plotted is the number of counts for the SUSY signal per 30 GeV and the square root of the total SM SNOWMASS backgrounds. The analysis is done at 200 fb$^{-1}$ of integrated luminosity, which gives a 5$\sigma$ discovery in this signal region. Right panel: The same analysis as in the left panel but for $E^{\text{miss}}_T$ with counts for the SUSY signal per 30 GeV.}
\label{fig:figure7}
\end{figure}

\begin{figure}[htp]
\centering
\includegraphics[width=.4\textwidth]{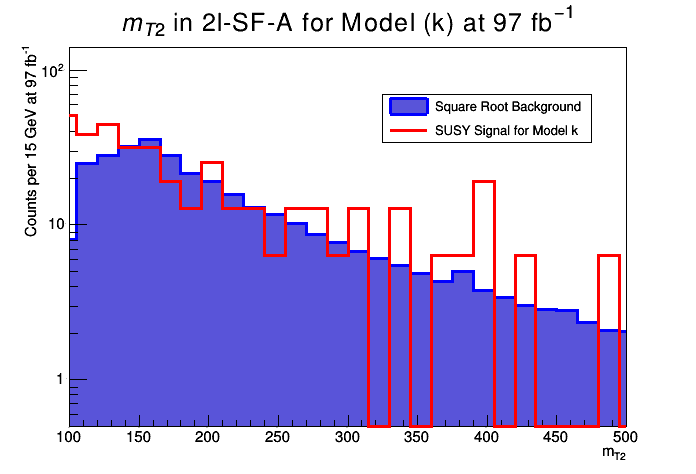}
\includegraphics[width=.4\textwidth]{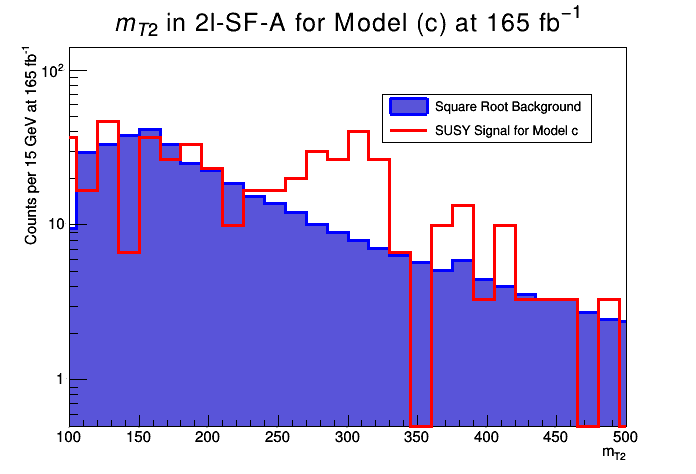}
\caption{Left panel: Distribution in $m_{\rm T2}$ for the 2l-SF-A signal region defined in Table~\ref{tab:SR2l} for stau coannihilation model (k) after cuts in that region. Plotted is the number of counts for the SUSY signal per 15 GeV and the square root of the total standard model SNOWMASS background. The analysis is done at 97 fb$^{-1}$ of integrated luminosity, which gives a 5$\sigma$ discovery in this signal region. Right panel: The same analysis as in the left panel but for model (c) at 165$\ifb$.}
\label{MT2_1}
\end{figure}

\begin{figure}[htp]
\centering
\includegraphics[width=.4\textwidth]{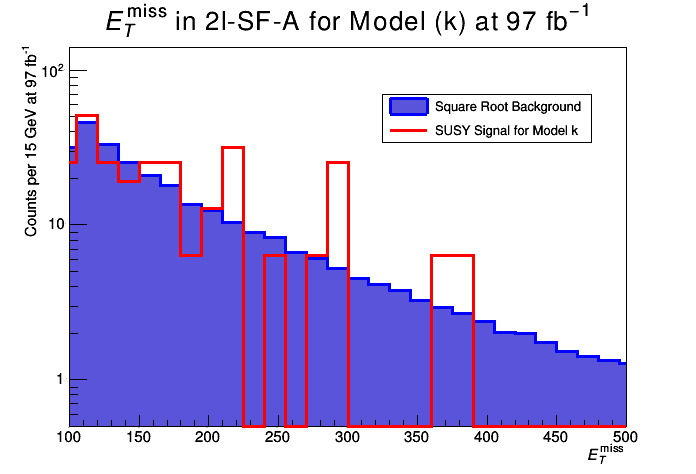}
\includegraphics[width=.4\textwidth]{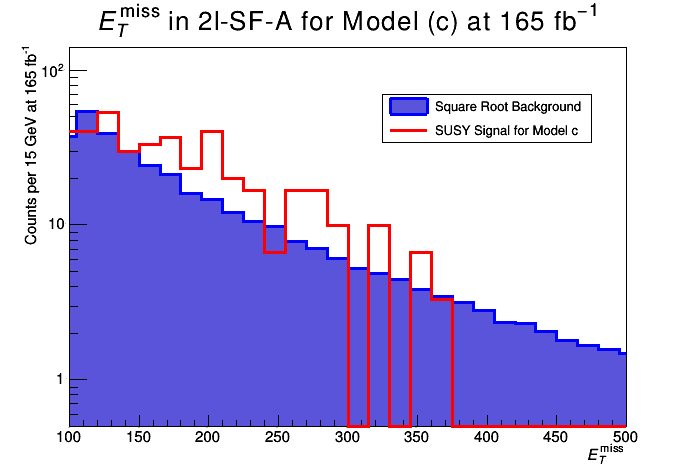}
\caption{Left panel: Distribution in $E_T^{\text{miss}}$ for the 2l-SF-A signal region defined in Table~\ref{tab:SR2l} for stau coannihilation model (k) after cuts in that region. Plotted is the number of counts for the SUSY signal per 15 GeV and the square root of the total standard model SNOWMASS background. The analysis is done at 97 fb$^{-1}$ of integrated luminosity, which gives a 5$\sigma$ discovery in this signal region. Right panel: The same analysis as in the left panel but for model (c) at 165$\ifb$.}
\label{MT2_2}
\end{figure}

\begin{figure}[H]
\begin{center}
	\includegraphics[width=0.45\textwidth]{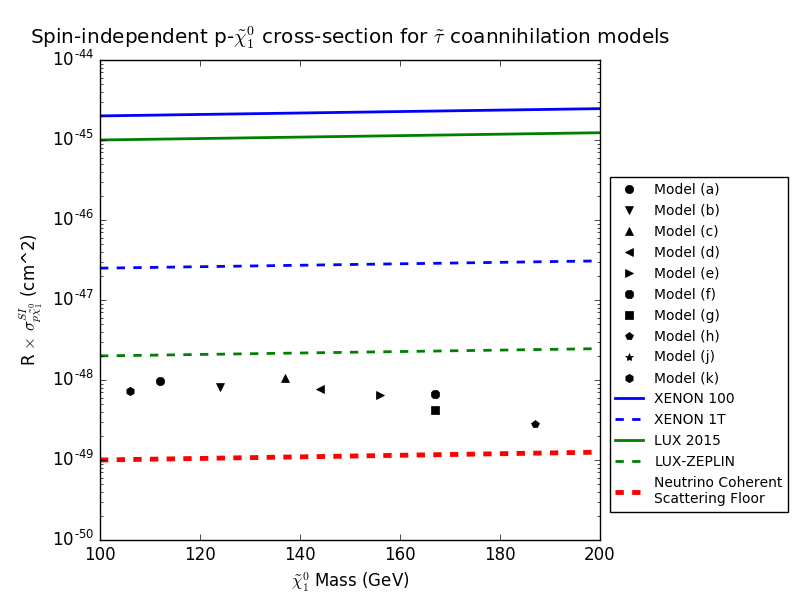}
	\includegraphics[width=0.45\textwidth]{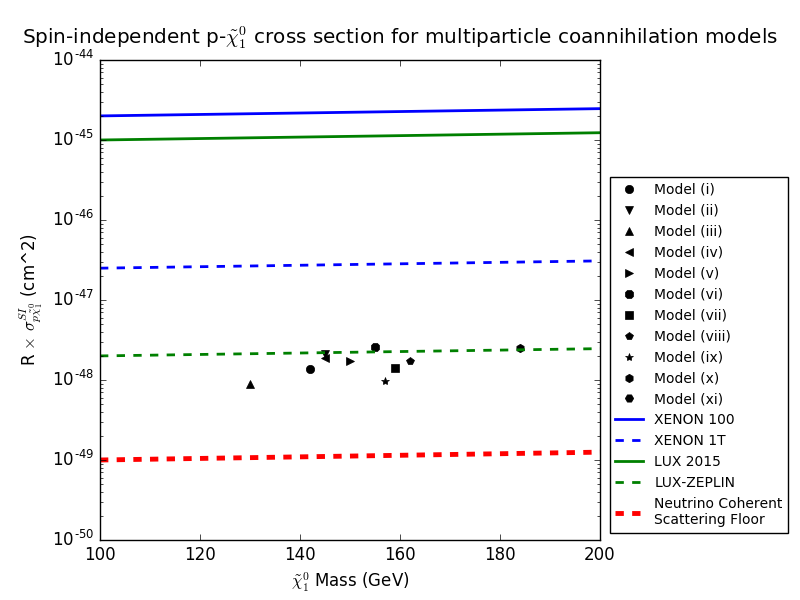}
\end{center}
    \caption{$R\times\sigma^{\text{SI}}_{p,\chi^0_1}$ for models of Tables 1 (left panel)  and Table 3 (right panel)
    as a function of LSP mass displayed alongside the current and projected range of the XENON and LUX experiments and the neutrino floor. }
    \end{figure}

\end{document}